%% file: MainDocument.tex
\documentclass[twocolumn]{aastex631}

\usepackage{longtable}
\usepackage{threeparttable}
\usepackage{booktabs}
\usepackage{longtable, threeparttablex}
\usepackage{float}

\shortauthors{Henry et al.}
\graphicspath{{./}{figures/}}

\begin{document}

\newcommand{\GAH}[1]{\textcolor{blue}{#1}}
\newcommand{\TvH}[1]{\textcolor{green}{#1}}

\title{Precise Stellar Age Constraints and Habitable Zone Evolution in Exoplanet Systems}

\author[0009-0005-9786-0166]{Grace A. Henry}
\affiliation{Department of Physical Sciences, Embry-Riddle Aeronautical University, 1 Aerospace Blvd, Daytona Beach, FL 32114, USA}

\author[0000-0002-5775-2866]{Ted von Hippel}
\affiliation{Department of Physical Sciences, Embry-Riddle Aeronautical University, 1 Aerospace Blvd, Daytona Beach, FL 32114, USA}

\author[0000-0002-9343-8612]{Anna Childs}
\affiliation{Center for Interdisciplinary Exploration and Research in Astrophysics (CIERA) and Department of Physics and Astronomy, Northwestern University}
\affiliation{Department of Physics and Astronomy, The University of Alabama, Box~870324, Tuscaloosa, AL~35487-0324, USA}



\begin{abstract}

Accurate stellar ages are fundamental to interpreting the evolutionary histories of habitable zone (HZ) planets. Because stellar luminosity evolves over time, HZ boundaries migrate outward, meaning that present-day HZ planets may have experienced different irradiation environments earlier in their evolution. Accurate ages are therefore required to estimate HZ residence times and place planetary habitability in an evolutionary context. We use the Bayesian Analysis of Stellar Evolution with nine parameters (BASE-9) code to constrain ages and masses for exoplanet host stars with planets located in the HZ. Our sample includes single stars with effective temperatures of 4475-7200 K that host planets within their calculated HZ boundaries. We fit isochrones to broadband photometry from Gaia DR3, Pan-STARRS, and 2MASS jointly with Gaia parallaxes, spectroscopic metallicities, and extinction estimates, and assess the resulting age posterior distributions. Using these age constraints, we track HZ migration along stellar evolutionary sequences to estimate continuous HZ residence times, allowing us to identify ideal targets for followup high resolution characterization. Because isochrones separate most strongly near and beyond the main-sequence turnoff, our analysis emphasizes evolved and turnoff hosts, where BASE-9 produces the tightest constraints. We quantify age precision across the sample, yielding highly precise ages for subgiants and evolved main-sequence or turnoff stars, and less precise but measurable ages for lower main-sequence hosts. The final catalog provides our age constraints for 149 HZ planets around 146 host stars and supports evolutionary interpretations of planetary histories.

\end{abstract}




\section{Introduction} \label{sec:intro}

The discovery of exoplanets beyond our Solar System has changed the landscape of planetary science and our understanding of where life might exist in the Universe. In just a few decades, exoplanet science has shifted from speculation to one of the fastest-growing fields in astrophysics. There are now thousands of confirmed exoplanets and an estimated billions more in our galaxy alone \citep{archive, Cassan2012}. Although our knowledge of planet-occurrence rates has advanced over the last few decades, the probability of life on another planet remains unknown.

A key piece of that puzzle is the habitable zone (HZ): the region around a star where an Earth-like planet could maintain liquid water on its surface \citep{Kasting1993}. Because the luminosity and effective temperature of a stara evolve, the HZ generally migrates outward with time, so habitability is not static \citep{Barnes2008, Kopparapu2013, Kopparapu2014}. Accurate stellar ages, combined with stellar evolution models, allow us to reconstruct the irradiation and activity history that planets experience and to interpret dynamical, atmospheric, and interior processes that unfold over Gyr timescales—from early migration and scattering to atmospheric escape and secondary atmosphere formation to the waning of volcanism and mantle convection \citep{Chatterjee2008, Nesvorny2018, Raymond2020, Juric2008, Raymond2022, Luger2015, Owen2012, Lopez2014, Kite2019, Foley2018, Noack2017, Ribas2005}. Age also guides biosignature expectations: younger systems may only support simple or transient signals, whereas older systems may have had time to develop more complex biospheres or even technosignatures \citep{Catling2020, Luger2015, Arney2016, LinLoeb2015, Wright2020, Sheikh2023}. In short, stellar age anchors planetary systems within their evolutionary context and helps assess whether worlds are entering, residing in, or exiting long-lived temperate conditions.

A variety of techniques exist to estimate stellar ages, each with strengths and limitations. Gyrochronology and activity indicators can be precise for middle-aged solar-type stars, but degrade for very young and very old regimes and depend on calibration samples \citep{Skumanich1972, Barnes2003, Barnes2007, Bouma2023}. Asteroseismology can produce tight constraints, but requires high-quality time-series photometry or spectroscopy and is available for a minority of targets \citep{Ulrich1986, Lebreton2014, Silva2015}. In contrast, isochrone-based age estimation compares observed parallaxes and multiband photometry to a grid of pre-computed isochrones. The fitting procedure accounts for uncertainties in metallicity and extinction and their covariance with age, allowing photometric measurements to be translated into calibrated stellar age and mass estimates \citep{Sandage1953, Eggen1962, Bertelli1994, Girardi2000}. By-eye fits, while intuitive and useful for quick checks, can vary between practitioners and may not consistently propagate uncertainties or account for correlations among parameters, making their age estimates less reproducible than those from probabilistic approaches. \citet{Bonfanti2015} applied a similar isochrone-based approach prior to the advent of Gaia data, providing an important reference for stellar age determination using earlier-generation photometry and models; we expand on this comparison in Section \ref{sec: litcompare}.

In this paper, we use BASE-9 (Bayesian Analysis of Stellar Evolution with nine parameters) to infer the ages of host stars with planets located in the HZ \citep{vonHippel2006}. BASE-9 performs fully Bayesian, self-consistent isochrone fitting across multiple photometric bands simultaneously, informed by priors on parallax, metallicity, and extinction, and returns posterior distributions for stellar age, metallicity, parallax, line-of-site extinction, and mass. We couple our BASE-9 stellar age and mass results with stellar evolutionary tracks to infer HZ planet properties and refine target selections for upcoming missions focused on the search for habitable worlds. Our goal is to demonstrate what multiband isochrone modeling can reveal about planetary system ages at scale. By producing robust, reproducible age estimates, we place HZ planets within their stellar evolutionary histories and set the foundation for evaluating the long-term stability of temperate conditions and the prospects for life.  In doing so, we can identify ideal targets for followup high resolution characterization.

\section{Methods} \label{sec:style}

In this section, we detail our methods for querying the \textit{NASA Exoplanet Archive Planetary Systems (ps)} table \citep{archive}, calculating the HZ zones for each system, interpreting our BASE-9 results, and finding the stellar evolutionary tracks which most closely represent the system's history. We restrict our sample to F-, G-, and K-type stars as these stars are long-lived enough to host life and usually bright enough to have quality photometry available. While Earth-sized planets are prioritized in the search for life, moons around larger planets could also harbor life. Thus, we consider all planet types with the only criterion being the planets reside in their system's HZ. 

\subsection{Data Selection from the NASA Exoplanet Archive}

The criteria for selecting stars are based on their effective temperature, mass, and availability of photometric data across multiple surveys. If the star's \textit{ps} default stellar parameters did not contain the needed values, we used the non-default values where available for the necessary data. If the data were not present after analyzing the non-default rows, we estimate the required parameters as we detail in Section \ref{sec:calc_missing_values}. Our original query of the \textit{NASA Exoplanet Archive} resulted in 165 planets around 161 stars, but two stars (GJ 2056 and HIP 34222) were later cut due to lack of sufficient prior metallicity data.

\subsubsection{Calculating Sample Selection Values} \label{sec:calc_missing_values}

To determine our initial sample selection, we only considered the single stars in the \textit{ps} table that have either an $T_{\text{eff}}$ or stellar mass measurement. If a star's $T_{\text{eff}}$ was missing, we estimated it using the mass-luminosity relation. If there was no luminosity value listed for a star, we calculated it in solar units using Stephan Boltzmann's equation, assuming there was a listed radius in solar units. If luminosity and radius were not provided in the \textit{ps} table, we calculated the radius in solar units using the $M_{\text{star}}$ value in solar mass units and assumed a solar density. This is followed by Stephen Boltzmann's law to calculate the final luminosity of the star. If the semi-major axis was missing from the \textit{ps} table, we estimated it using the observed planet orbital period and the simplified form of Kepler's third law.  Lastly, we select stars with observed or approximated $4475 \leq T_{\text{eff}} \leq 7200$ K to ensure the inclusion of F-, G-, and K-type main-sequence stars. We emphasize that approximate stellar parameters were used only for sample selection and do not affect the BASE-9 results in any way.

\subsection{Calculating the Inner and Outer HZ Boundaries}

To determine the HZ boundaries for each system, we used the updated empirical formulation from \citet{Kopparapu2013}. The method is based on a 1D radiative-convective climate model that estimates the effective stellar flux ($S_{\mathrm{eff}}$) required to maintain surface liquid water on an Earth-like planet. The distance of the HZ boundaries from the star is then calculated using the following.

\begin{equation}
    d = \left( \frac{L_{\star}/L_{\odot}}{S_{\mathrm{eff}}} \right)^{1/2}
\end{equation}

where $d$ is the orbital distance in AU, $L_{\star}$ is the stellar luminosity, and $S_{\mathrm{eff}}$ is the effective stellar flux received at the top of the planetary atmosphere, which depends on the stellar effective temperature ($T_{\mathrm{eff}}$). The $S_{\mathrm{eff}}$ for each HZ limit is approximated by:

\begin{equation}
    S_{\mathrm{eff}} = S_{\mathrm{eff},\odot} + a T_{\star} + b T_{\star}^2 + c T_{\star}^3 + d T_{\star}^4
\end{equation}

where $T_{\star} = T_{\mathrm{eff}} - 5780$ K, and the coefficients $a$, $b$, $c$, and $d$ differ depending on the HZ limit. Coefficient values define different HZ boundaries: Recent Venus, corresponding to the inner edge where a Venus-like planet is inferred to have undergone a runaway greenhouse; Runaway Greenhouse, marking the onset of rapid ocean loss; Maximum Greenhouse, defining the outer edge where the greenhouse warming of CO$_2$ is maximized; and Early Mars, representing the outer limit consistent with liquid water on early Mars. The solar-normalized flux $S_{\mathrm{eff},\odot}$ and the corresponding coefficients are given in \citet{Kopparapu2013} Table~3.

For the optimistic HZ, we use the inner boundary defined by the Recent Venus limit and the outer boundary by the Early Mars limit. These Recent Venus and Early Mars limits are the boundaries adopted in the subsequent HZ migration analysis (Section \ref{sec:Implications_of_HZ}, below).

We adopt these limits because they represent the empirical outermost constraints on the inner and outer edges of habitability derived from Solar System analogs. Using these broader boundaries allows us to include systems that may plausibly sustain surface liquid water under a wider range of atmospheric and geophysical conditions. Given the many uncertainties that influence planetary habitability, we intentionally avoid imposing stricter theoretical limits (e.g., Runaway or Maximum Greenhouse) that could prematurely exclude potentially habitable environments. This approach yields a more inclusive sample while still remaining grounded in physically motivated limits.

If, after these calculations, a planet is determined to reside within the optimistic HZ of the system, we include its host star in our analysis. In multi-planet systems, each planet is evaluated individually with respect to the HZ boundaries, and if multiple planets fall within the HZ, all such planets are included in the analysis.

\subsection{Photometric Data Query \& Prior Selection}

To derive stellar ages with BASE-9 we need extinction, metallicity, and distance priors in addition to high-precision photometry for each star. Because the \textit{ps} table only provides common names for the stars, we first cross-match the common names to the Gaia DR3 source ID with SIMBAD. Next, we queried Gaia DR3, Pan-STARRS (Panoramic Survey Telescope and Rapid Response System), and 2MASS (Two Micro All Sky Survey) data using Gaia's advanced search feature \citep{Gaia, Panstarrs, 2Mass}. We chose these surveys because of their high-precision and multi-wavelength coverage, crucial for accurately characterizing stellar properties.

This query also included measurements of Gaia DR3 parallax and extinction ($A_{G}$). The Gaia DR3 parallaxes have a mean signal-to-noise ratio $\langle \mathrm{S/N}\rangle_{\varpi}=752$, with a minimum of $(\mathrm{S/N})_{\varpi,\min}=\text{18.5}$. If there was no value available for $A_{G}$, we assumed a value of 0. However, we note that all missing values $A_{G}$ were for stars that reside within $100 \, \rm pc$, thus making our assumption of 0 reasonable.

\begin{figure}[!t]
    \centering
    \includegraphics[width=0.82\columnwidth]{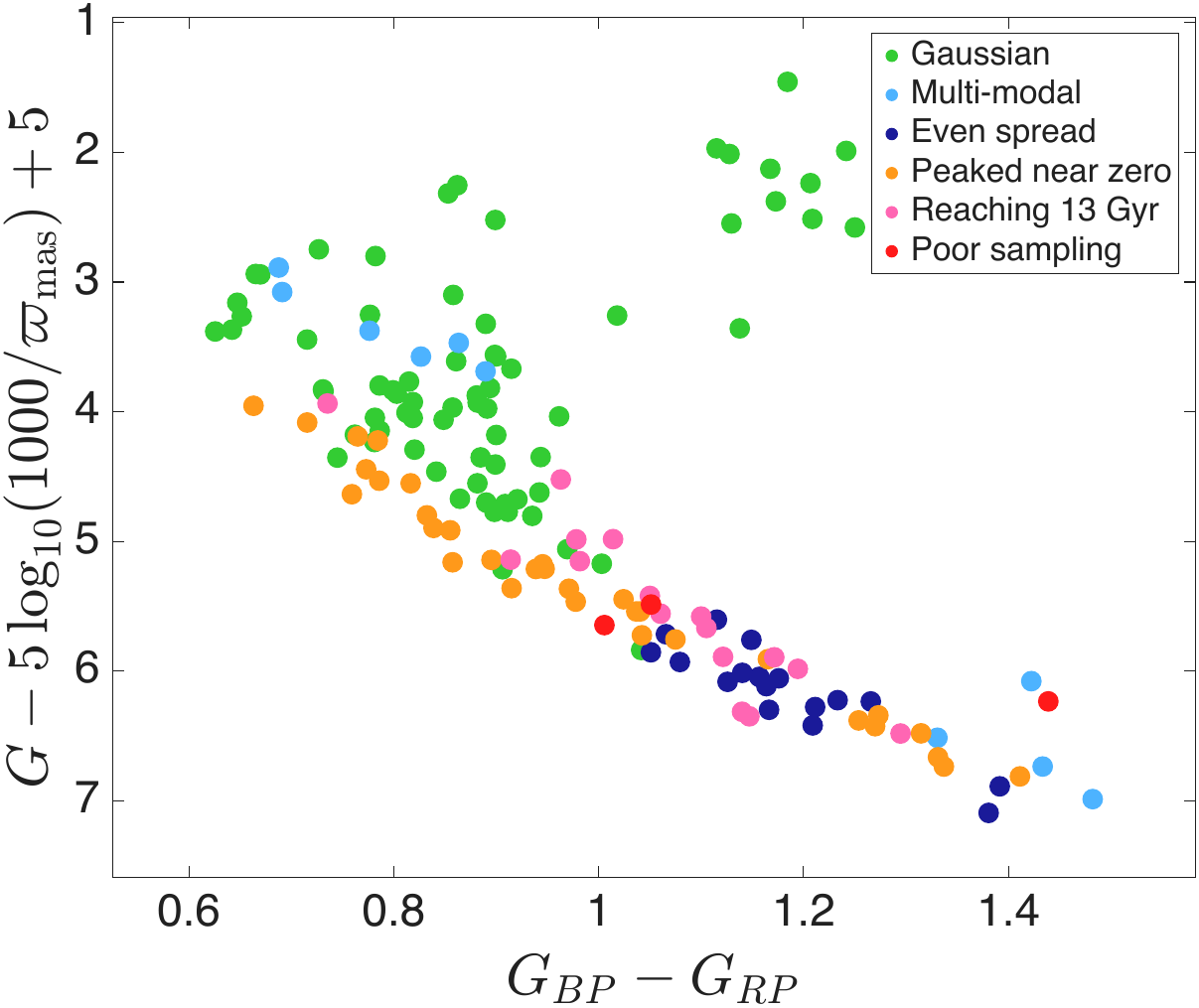}

    \vspace{0.25em}

    \includegraphics[width=0.82\columnwidth]{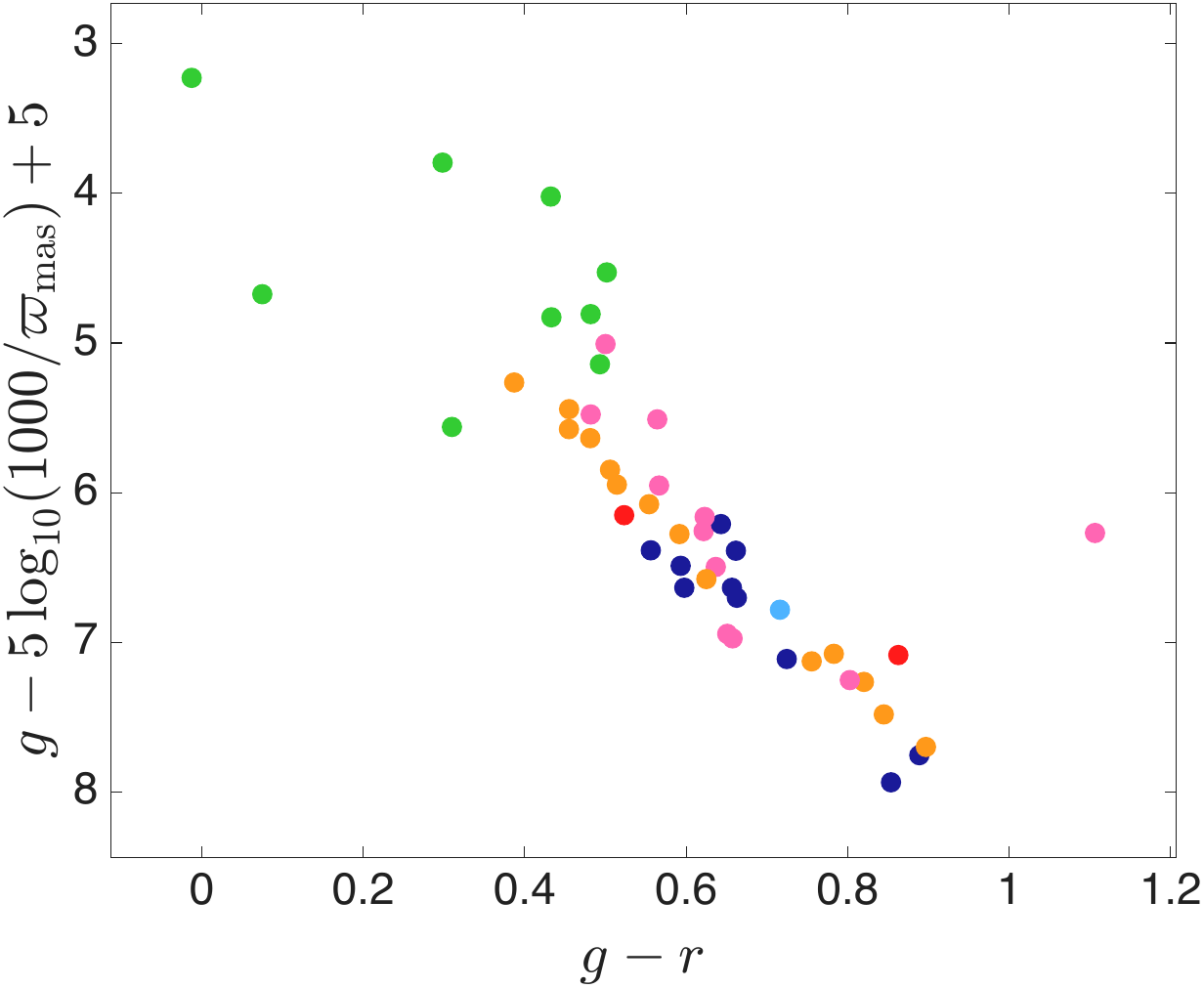}

    \vspace{0.25em}

    \includegraphics[width=0.82\columnwidth]{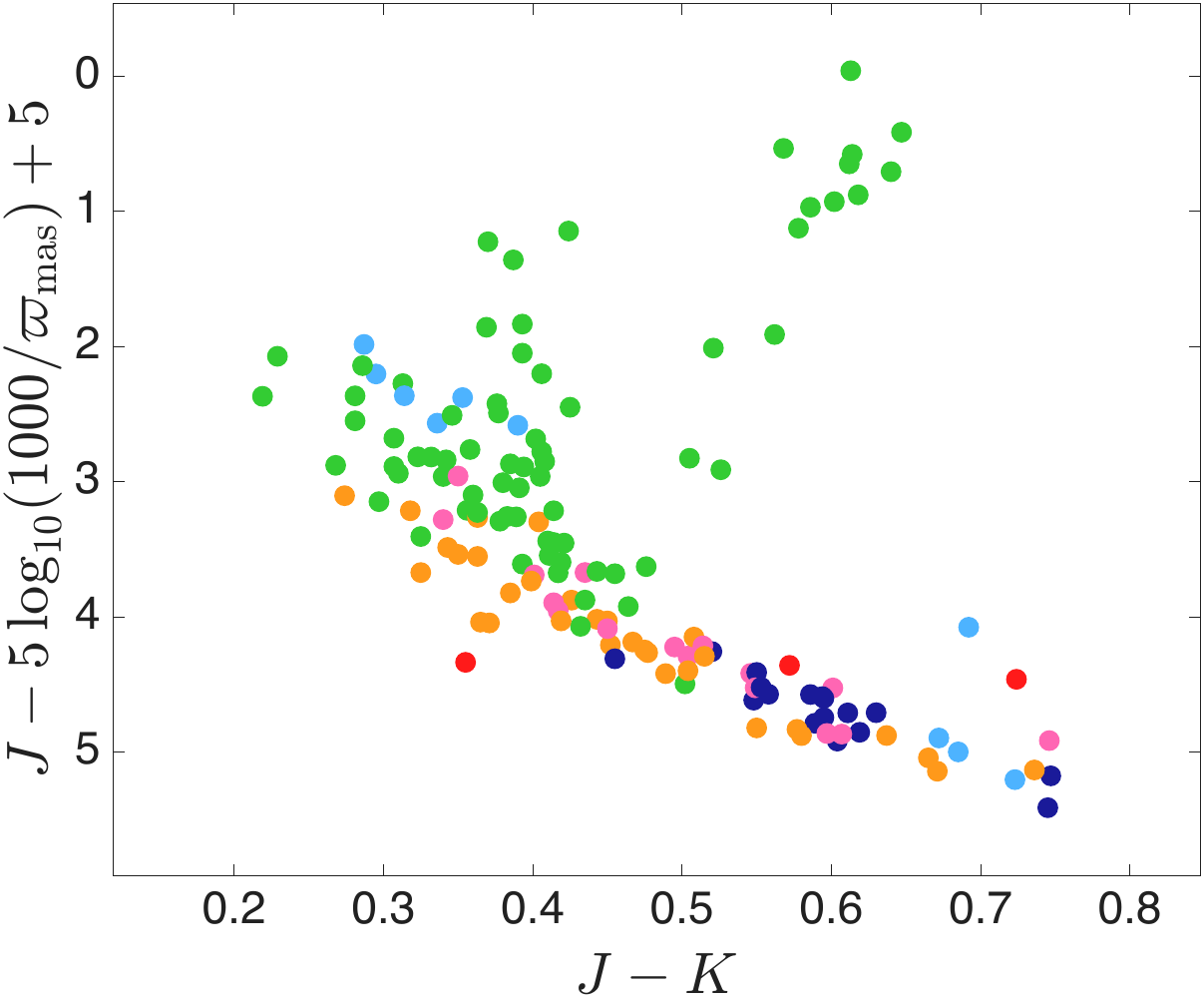}

    \caption{Gaia, Pan-STARRS, and 2MASS color-magnitude diagrams, respectively, for the stars evaluated in this study.}
    \label{fig:CMDs}
\end{figure}

Although Gaia DR3 photometry $G$, $G_{BP}$, and $G_{RP}$ is available for all stars, not all stars have g, r, i, z, and y photometry from Pan-STARRS or J, H, and K photometry from 2MASS. This is especially true for stars south of $-30^{\circ}$ where Pan-STARRS did not survey. Although 2MASS is a full-sky survey, some stars do not have 2MASS photometry because they have a K band magnitude outside of the -4 to 16 magnitude range. To ensure only reliable photometry was used, we applied the recommended saturation limits for each survey. Pan-STARRS photometry brighter than the survey's conservative saturation limits ($g=14.5$, $r=15$, $i=15$, $z=14$, and $y=13$ mag) was excluded, while the remaining unsaturated Pan-STARRS bands were retained. For 2MASS, stars with $K \lesssim 3$ mag and poor photometric quality flags indicative of saturation (47 Uma, HD 10647, HD 10697, HD 154391, HD 160691, HD 20794, HD 30562, HD 69830, HIP 67851, and tau Cet) were excluded from the analysis because only Gaia and unreliable 2MASS photometry were available for these systems, whereas our analysis requires photometry from at least two independent surveys. Figure \ref{fig:CMDs} shows Gaia DR3, Pan-STARRS and 2MASS color-magnitude diagrams (CMDs) for the stars where the photometry in these surveys is available. The CMDs include color codes for each star based on its posterior sampling classification (see Section \ref{sec:Results}.)

To obtain stellar metallicity priors, we use SweetCAT \citep{Sousa2021}. If metallicity values were not available in SweetCAT for a given star, we took metallicity values from the \textit{ps} table. Table \ref{tab:priors} lists the prior values used for each star, as well as the reference for each [Fe/H] value.

\subsection{Posterior Distributions with BASE-9}

To constrain the ages and masses of our target stars, we used the BASE-9 software suite, a Markov Chain Monte Carlo (MCMC) tool designed to derive stellar parameters from photometric and astrometric data \citep{vonHippel2006, vanDyk2009, Stenning2016}. BASE-9 compares observational measurements to theoretical stellar evolution models within a Bayesian framework, producing posterior probability distributions that reflect both the data and prior assumptions.

The core inputs to BASE-9 include photometric magnitudes across multiple filters, parallax measurements (from Gaia DR3), metallicity estimates ([Fe/H]) and extinction values (also from Gaia DR3). These parameters constrain the star’s position on the CMD and allow the software to fit stellar evolution models, specifically for this study PARSEC isochrones \citep{Bressan2012}, to the data. BASE-9 samples the joint posterior probability space using MCMC to yield distributions for age, metallicity, parallax, and extinction, with age as the primary parameter of interest in this study.

In addition to age estimation, we used the BASE-9 \textit{sampleMass} module to derive posterior mass distributions for each star. This module samples from the MCMC outputs on age, distance, metallicity, and absorption and compares them to the same stellar models to infer a physically consistent stellar mass for each MCMC iteration. Since stellar age is closely coupled to mass via evolutionary timescales, incorporating posterior mass estimates provides a reliability check of BASE-9's age determinations. These values are additionally useful for refining dynamical studies of planetary systems.

Using this probabilistic approach, our analysis accounts for measurement uncertainties and parameter degeneracies, providing robust estimates of the age and mass of each star. BASE-9 returns a posterior distribution of parameters for each star. For compactness, we take the median and $16^{\rm th}$ and $84^{\rm th}$ percentile values of the posterior distribution for the stellar age and its uncertainty. These summary statistics are then used to investigate the evolutionary states of host stars and their planetary systems.

\subsection{Stellar Evolutionary Tracks}

BASE-9 does not directly produce a continuous stellar evolutionary track for an individual star. Therefore, to trace the luminosity and effective temperature evolution of each host star through time, we use the PARSEC v1.2 stellar evolutionary tracks for non-rotating stars from \cite{Bressan2012, Chen2014, Chen2015, Fu2018} \footnote{All PARSEC v1.2 stellar evolutionary tracks were downloaded from \url{https://stev.oapd.inaf.it/PARSEC/tracks_v12s.html}}. The BASE-9 posterior metallicity, age, and mass estimates are used to identify the closest corresponding evolutionary track within the PARSEC grid. Specifically, we first find the set of tracks with the closest matching metallicity. Next, we compute a $\chi^2$ statistic between the BASE-9 posterior age and mass and every age and mass point along the set of closest matching metallicity PARSEC evolutionary tracks. The evolutionary track with the corresponding minimum $\chi^2$ is adopted for each system.

We begin tracking HZ evolution at 30 Myr, assuming that planet formation is largely complete following the dissipation of the protoplanetary disk. Observations of protoplanetary disks suggest that disk lifetimes rarely exceed $\sim$30 Myr, providing an empirical upper limit on the epoch by which planets are expected to have formed and begun to evolve independently of their natal disks \citep{Long2025}.

\section{Results}\label{sec:Results}

\subsection{Posterior Distribution Types} \label{sec:resulttype}

In this section, we present the results of our BASE-9 analyzes. Given the variety in posterior convergence and distribution shapes in the stellar sample, we first organize the results by the shape and behavior of each star’s posterior distribution. These types include well-converged nearly Gaussian posteriors, complex multimodal shapes, and non-converged outputs. By characterizing these distributions, we assess the reliability of each age estimate. We then compile the median ages and uncertainties in Table \ref{tab:results} before exploring the implications of these stellar ages for the exoplanets orbiting each host star.

\subsubsection{Nearly Gaussian Posterior Distributions}

\begin{figure}[!h]
  \centering
  \includegraphics[width=\columnwidth]{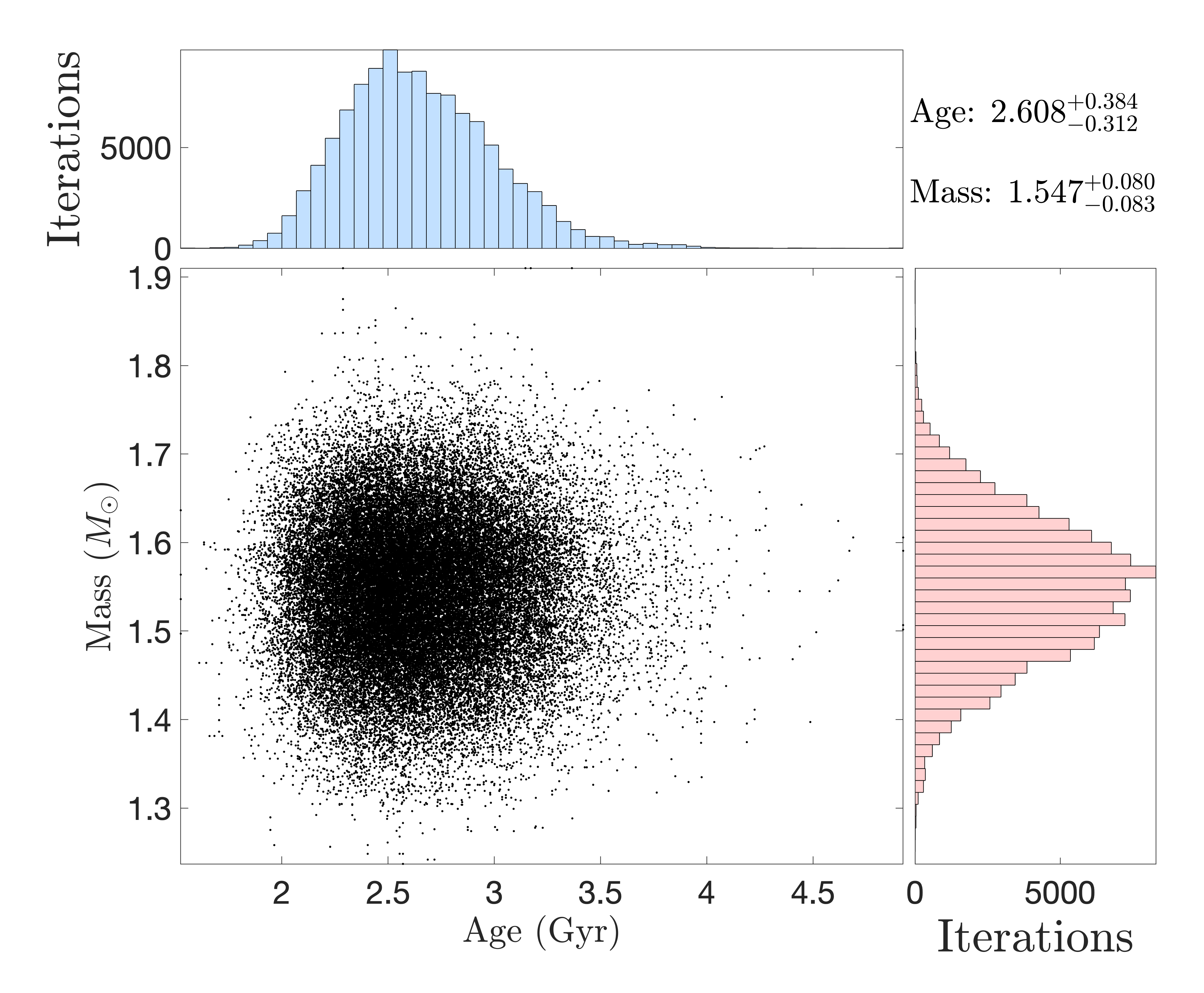}
  \caption{Posterior distribution of stellar mass versus age for HD 145934. The central scatter plot shows the joint distribution of age (in Gyr) and mass (in solar masses) across all sampled iterations. Marginal histograms along the top and right display the age and mass distributions, respectively, highlighting the parameter space explored during the MCMC sampling.}
  \label{fig:gaussian}
\end{figure}

Among the stars analyzed, several produced posterior distributions that were well approximated by a Gaussian profile. These MCMC chains converged quickly, and their distributions display a single nearly symmetric peak and narrow uncertainty intervals of 68\%. Figure \ref{fig:gaussian} shows a representative example of such a nearly Gaussian age posterior for the star HD 145934.

In these cases, the median age estimate is robust as the shape and behavior of the sampling chain indicate the parameter space was thoroughly explored and the posterior maximum was well-sampled. These stars also typically exhibited well-constrained posterior parallaxes and metallicities consistent with the priors, contributing to the model’s ability to localize the age with precision.

\subsubsection{Multimodal Posterior Distributions}

\begin{figure}[!h]
  \centering
  \includegraphics[width=\columnwidth]{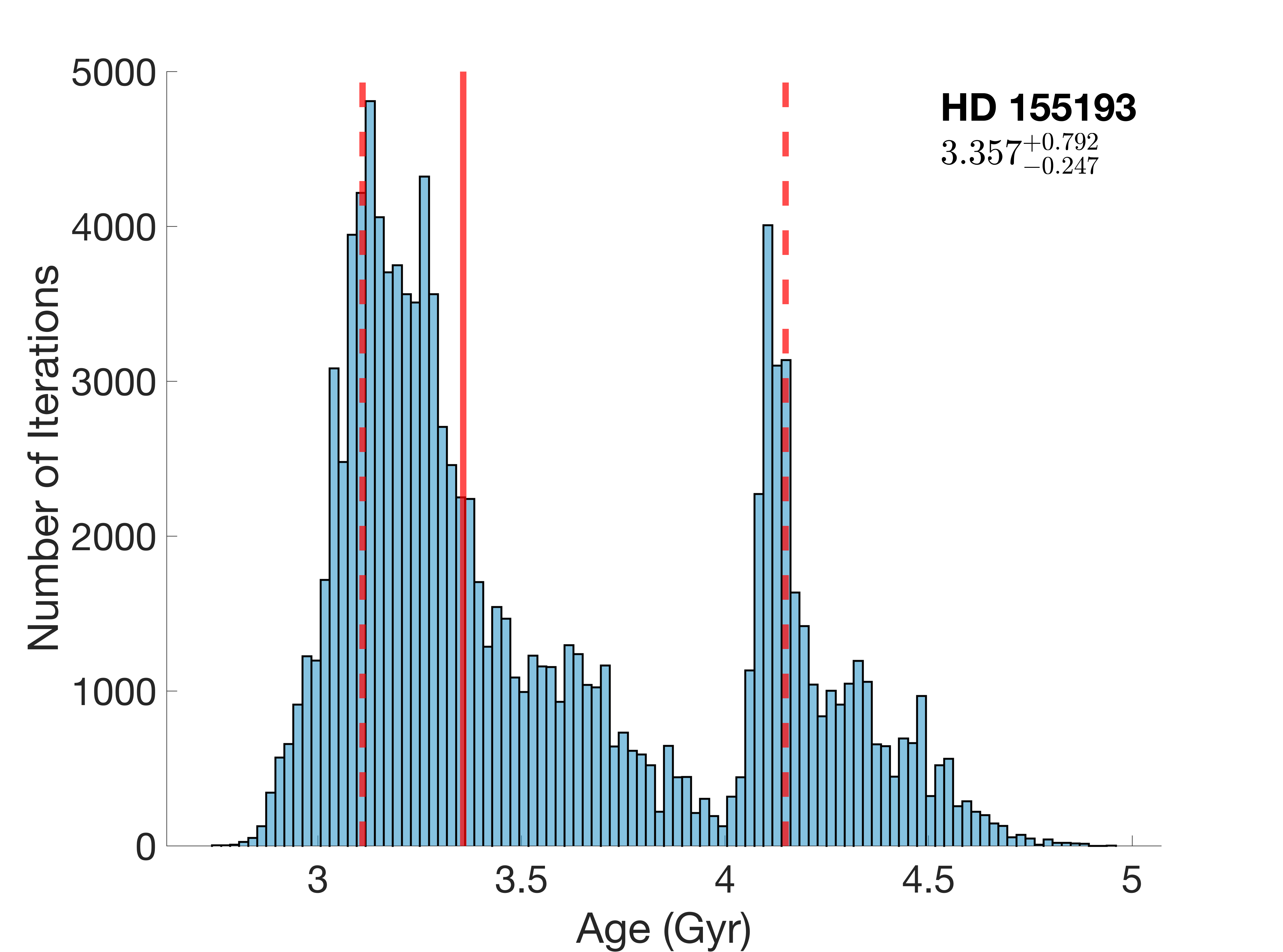}
  \caption{Posterior age distribution for HD 155193 showing multiple peaks. The solid red line marks the median; dashed lines indicate the 16th and 84th percentiles.}
  \label{fig:multimodal}
\end{figure}

For several stars, the age posterior distributions are multimodal. Figure \ref{fig:multimodal} shows an example for HD 155193. Such multimodal posteriors typically indicate that several distinct evolutionary stages are statistically compatible with the available photometric and astrometric data. This ambiguity often arises near isochrone degeneracies, where stars of different ages can share similar observable properties.

As illustrated in Figure \ref{fig:CMDs} stars with multimodal age posteriors cluster near the main-sequence turnoff and early subgiant branch, where isochrones of different ages converge in color–magnitude space. In this region of the CMD, distinct evolutionary phases project onto similar, more evolved tracks and overlap younger ones, so the same photometric position is consistent with multiple stellar ages. This geometric crowding of isochrones naturally produces multiple peaks in the inferred age distribution.

\subsubsection{Non-convergence: Evenly Spread Posterior Distributions}

Some stellar posterior distributions show a flat or broadly uniform spread across the full age range of 0 to 13 Gyr. These cases indicate that BASE-9 was unable to converge on a specific age solution, and the prior data constraints were insufficient to break the degeneracies in the stellar evolution tracks. This pattern is often a sign that the star lies in a region of the CMD where evolutionary changes occur slowly, such as on the lower main-sequence, where age has little influence on observable properties.

\begin{figure}[!h]
  \centering
  \includegraphics[width=\columnwidth]{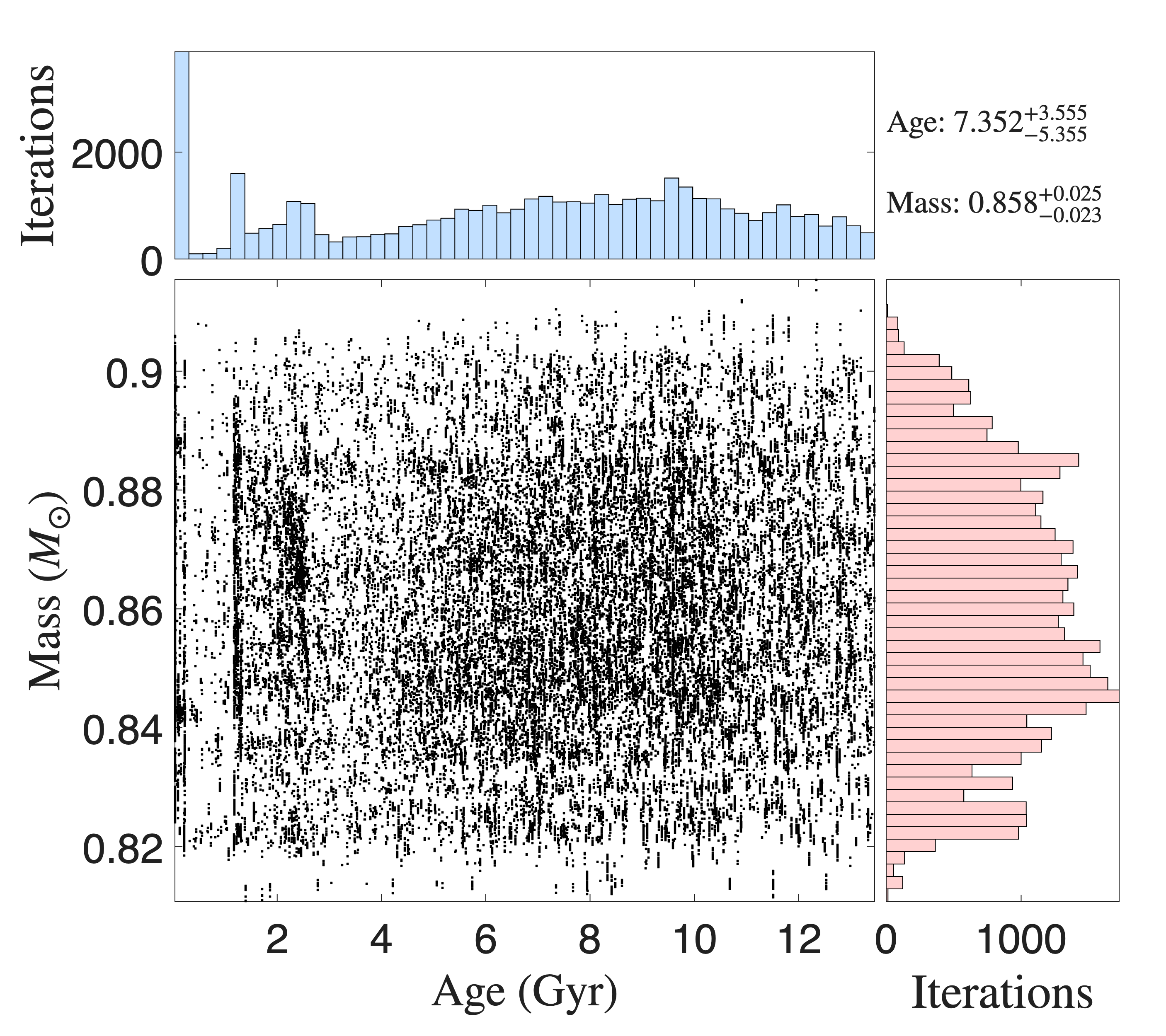}
  \caption{Mass vs. age joint posterior distribution for HD 114783.}
  \label{fig:flat}
\end{figure}

An example of this behavior is presented in Figure \ref{fig:flat} for HD 114783. The plot reveals a near-uniform distribution, suggesting that the star's photometry does not strongly align with any single evolutionary stage. Without stronger priors or substantially improved observational data, the model distributes probability roughly equally across all plausible ages, emphasizing the difficulty in dating such stars through photometric methods.

\subsubsection{Posterior Distributions Peaked near Zero}

\begin{figure}[!h]
  \centering
  \includegraphics[width=\columnwidth]{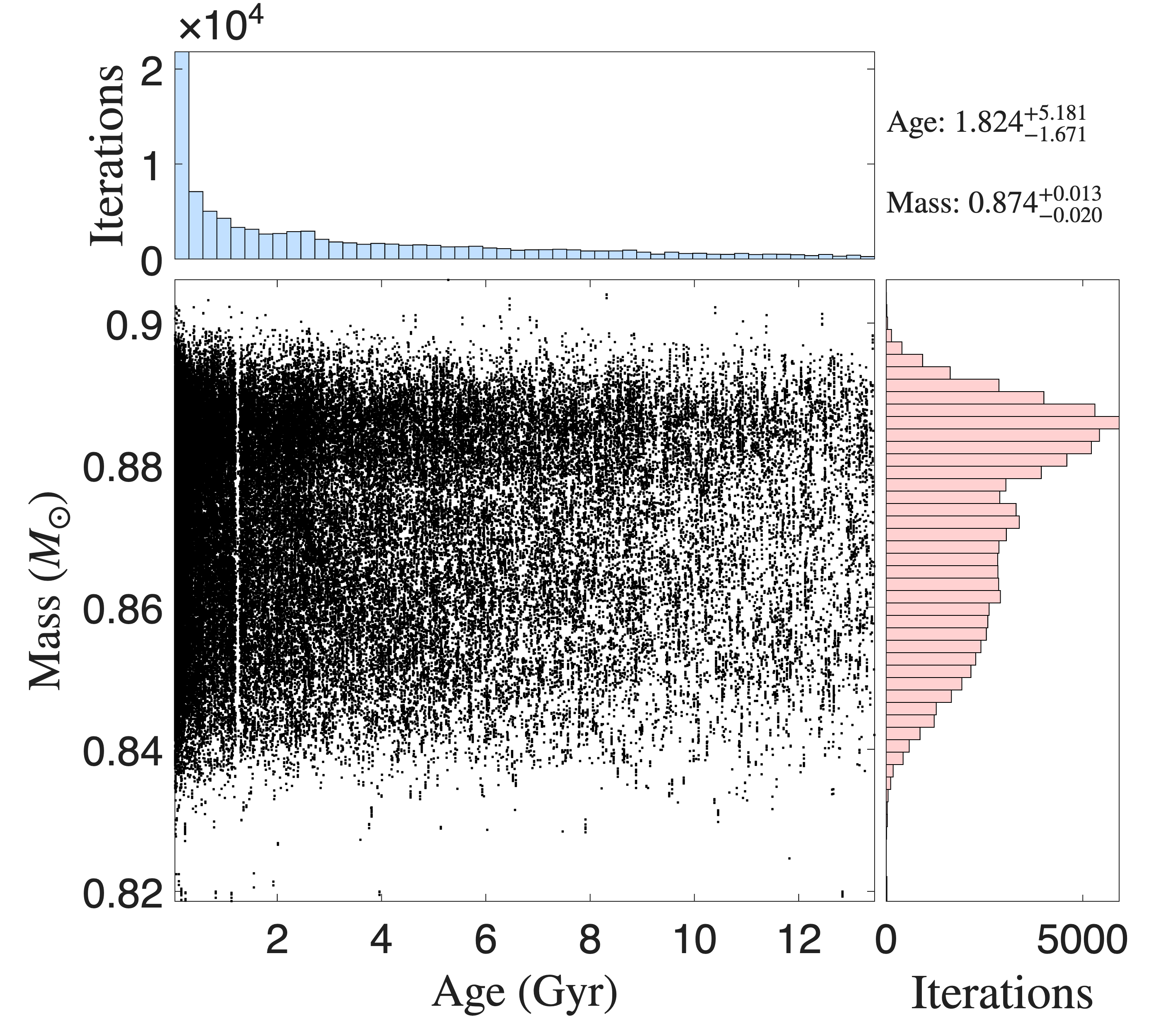}
  \caption{The distribution for HD 128356 demonstrating a most likely age near zero Gyr, but with a wide range of possibilities up to 3 or 4 Gyr.}
  \label{fig:zero}
\end{figure}

In several cases, the age posterior distributions peaked near zero, suggesting that the models favored extremely young stellar ages. An example of this kind of distribution is shown in Figure \ref{fig:zero} with HD 128356.

Stars with this behavior typically show strong peaks near zero Gyr with the probability decreasing as the models explore greater ages. Many of these stars are likely to be very young, though there is always the possibility that some of the photometry or the priors may be in error.

\subsubsection{Posterior Distributions Reaching 13 Gyr}

In contrast to stars whose posterior age distributions peak sharply near young ages, some stars exhibit posterior age distributions that extend to the model’s upper limit of 13 Gyr. These broad posteriors arise when the observed photometry and parallax place the star in a region of the CMD where isochrones are closely spaced at old ages.

\begin{figure}[!h]
    \centering
        \includegraphics[width=\columnwidth]{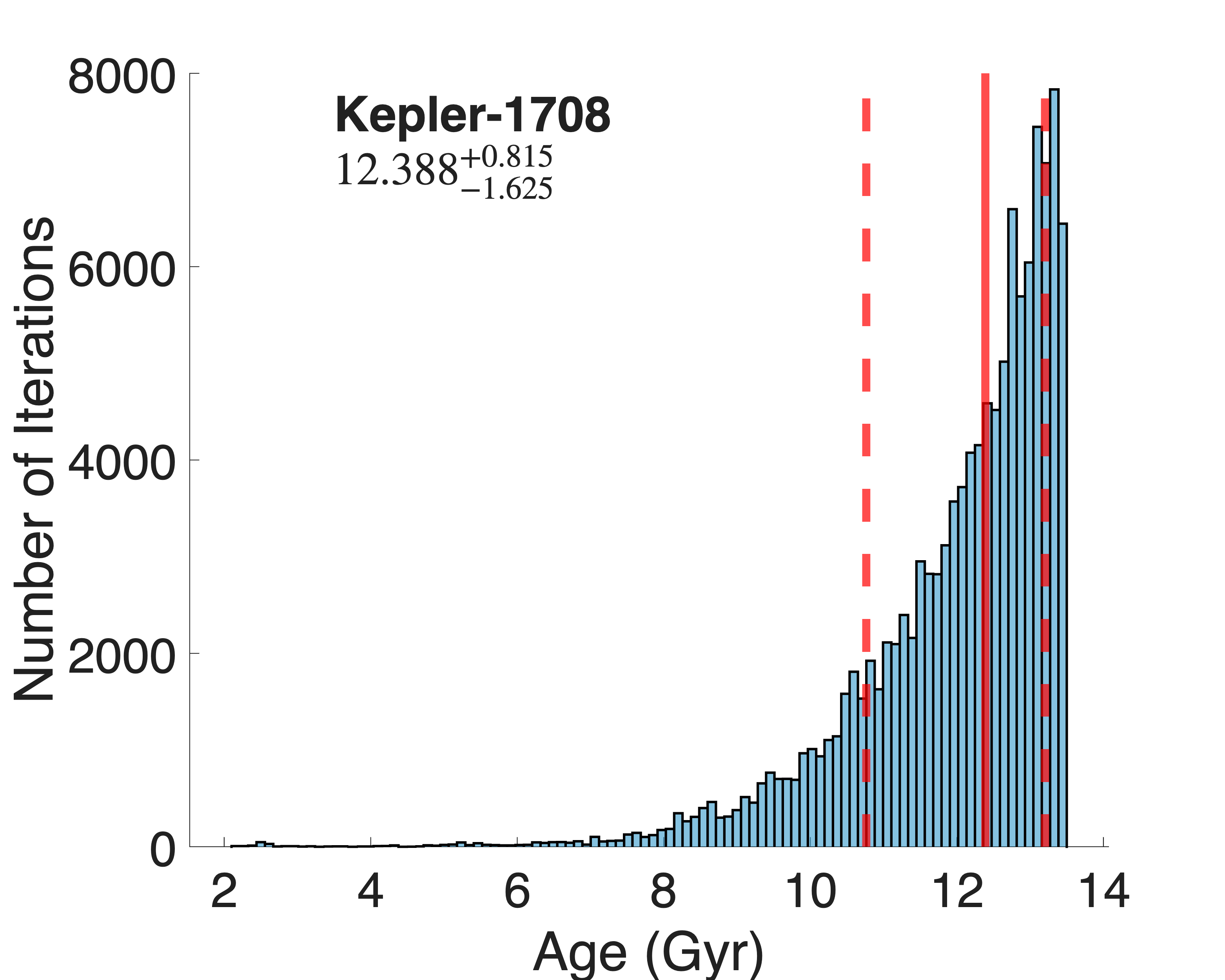}
        \caption{Posterior age distribution for Kepler-1708, demonstrating an age estimate approaching the model’s upper limit.}
    \label{Fig:13Gyr}
\end{figure}

Most of these stars lie on the cooler side of the main sequence rather than in the subgiant region (as seen in Figure \ref{fig:CMDs}). In this low-mass regime, slow stellar evolution yields posterior distributions with long, shallow tails that rise only as the MCMC sampler reaches the maximum allowed age. Figure \ref{Fig:13Gyr} shows this behavior for Kepler-1708.

This upper-limit behavior may reflect underestimated reddening, incorrect metallicity, binarity, or parallax uncertainties, all of which can shift the star’s CMD position and make it appear older than the oldest available isochrone.

\subsection{BASE-9 Posterior Distributions}

In this section are the final results from the MCMC algorithm used to derive stellar age, mass, and posterior values for metallicity, parallax, and dust extinction. The results are summarized in Table \ref{tab:results}. Special circumstances are explained in more detail in Section \ref{sec: Discussion}.

To synthesize the range of posterior distributions produced by our modeling, we assign each star an interpretation code that summarizes the qualitative behavior of its posteriors across five key parameters: age, metallicity ([Fe/H]), parallax, extinction (A$_{G}$), and stellar mass. Each digit in this five-character code corresponds to a distinct distribution type for those parameters, respectively.

The first digit, ranging from 1 through 5, classifies the age posterior by its shape. These types are consistent with those outlined in Section \ref{sec:resulttype}:

\begin{enumerate}
    \item Narrow Gaussian
    \item Multimodal
    \item Peaked near zero
    \item Reaching 13 Gyr
    \item Non-convergence
\end{enumerate}

The remaining four digits are binary flags for metallicity, parallax, dust extinction, and mass:

\begin{enumerate}
    \item The posterior distribution is consistent with the prior.
    \item The posterior deviates from the prior.
\end{enumerate}

This system provides a compact summary of the behavior of each posterior relative to its prior. For example, a code of 31212 indicates an age posterior of type 3 (peaked near zero), metallicity consistent with its prior, parallax deviating from its prior by more than $3\sigma$, consistent extinction, and stellar mass inconsistent with previously published estimates. For each parameter, consistency is evaluated by comparing the posterior median to the prior mean and flagging deviations exceeding three times the prior standard deviation. For stellar mass, this comparison is made against the values reported in the \textit{ps} table using the same $3\sigma$ criterion.

In general, when a posterior closely matches the prior this suggests that single star stellar evolution models are consistent with the priors and the photometric data and conversely when the priors for parallax, metallicity, and extinction are inconsistent with the posteriors for these quantities, single star stellar evolution models cannot fit them and the photometry simultaneously. Stars with strongly non-Gaussian age posteriors are not necessarily problematic, because stellar evolution is highly nonlinear.\footnote{To view the data files for each star, visit \url{https://doi.org/10.5281/zenodo.22212897}.}

\subsection{HZ Migration Along Stellar Evolutionary Tracks}

Applying our stellar age constraints to the time evolution of HZ boundaries reveals a diverse range of planetary residence histories across the sample. By comparing the present-day orbital distance of each planet with the evolving HZ limits, we identify clear differences in how long and under what conditions planets have occupied temperate environments.

\begin{figure*}[ht]
    \centering
    \includegraphics[width=\columnwidth]{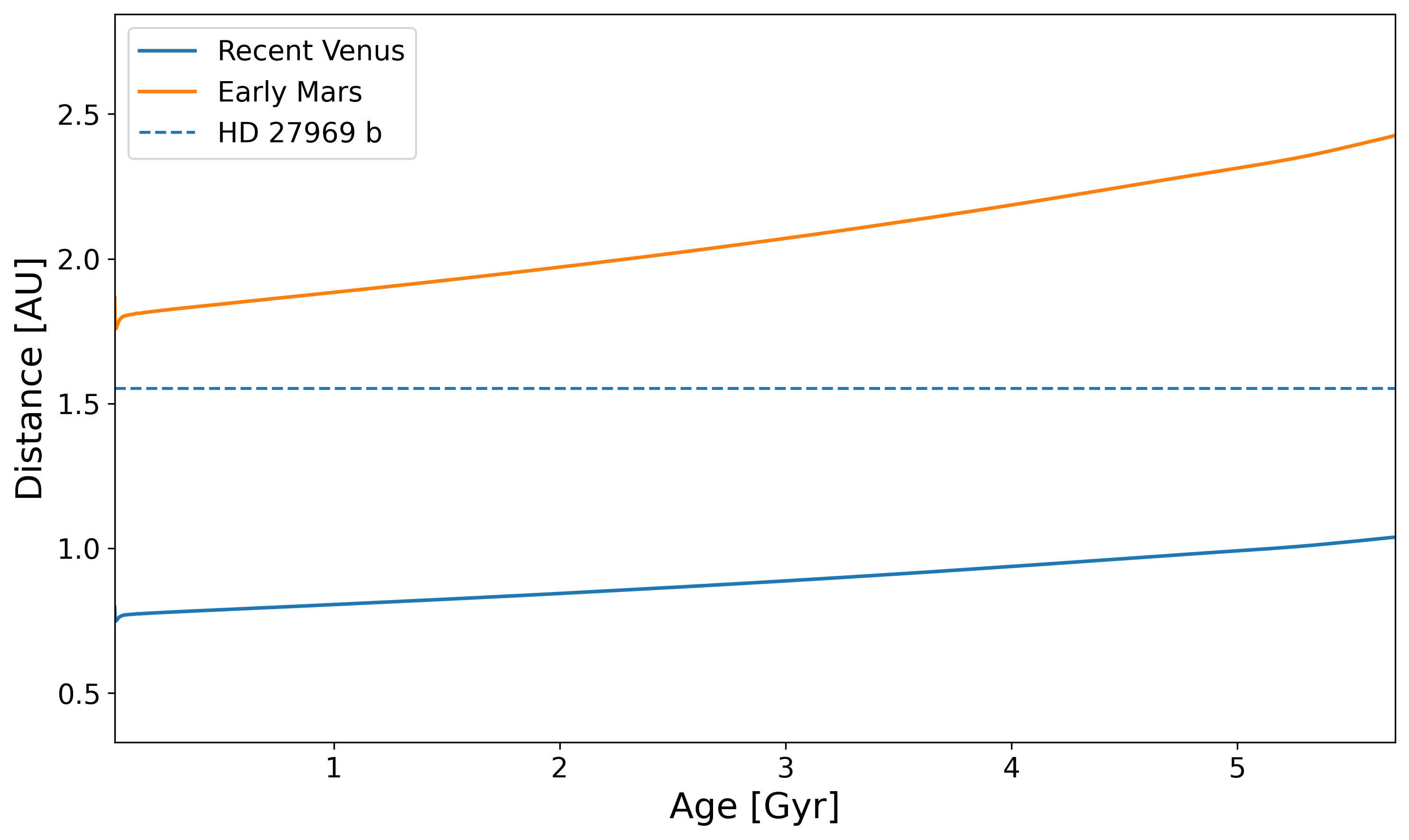}
    \includegraphics[width=\columnwidth]{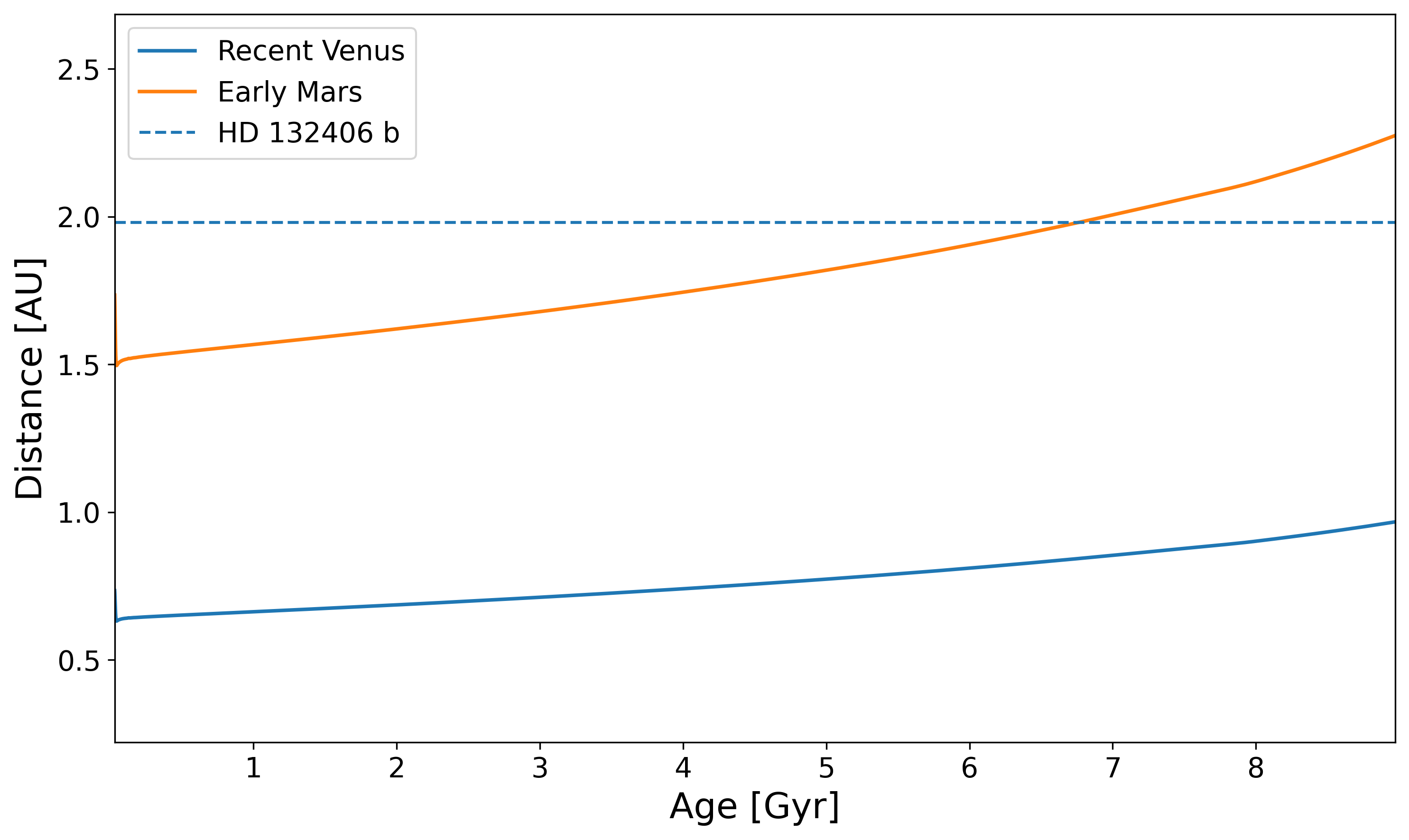}
    \includegraphics[width=\columnwidth]{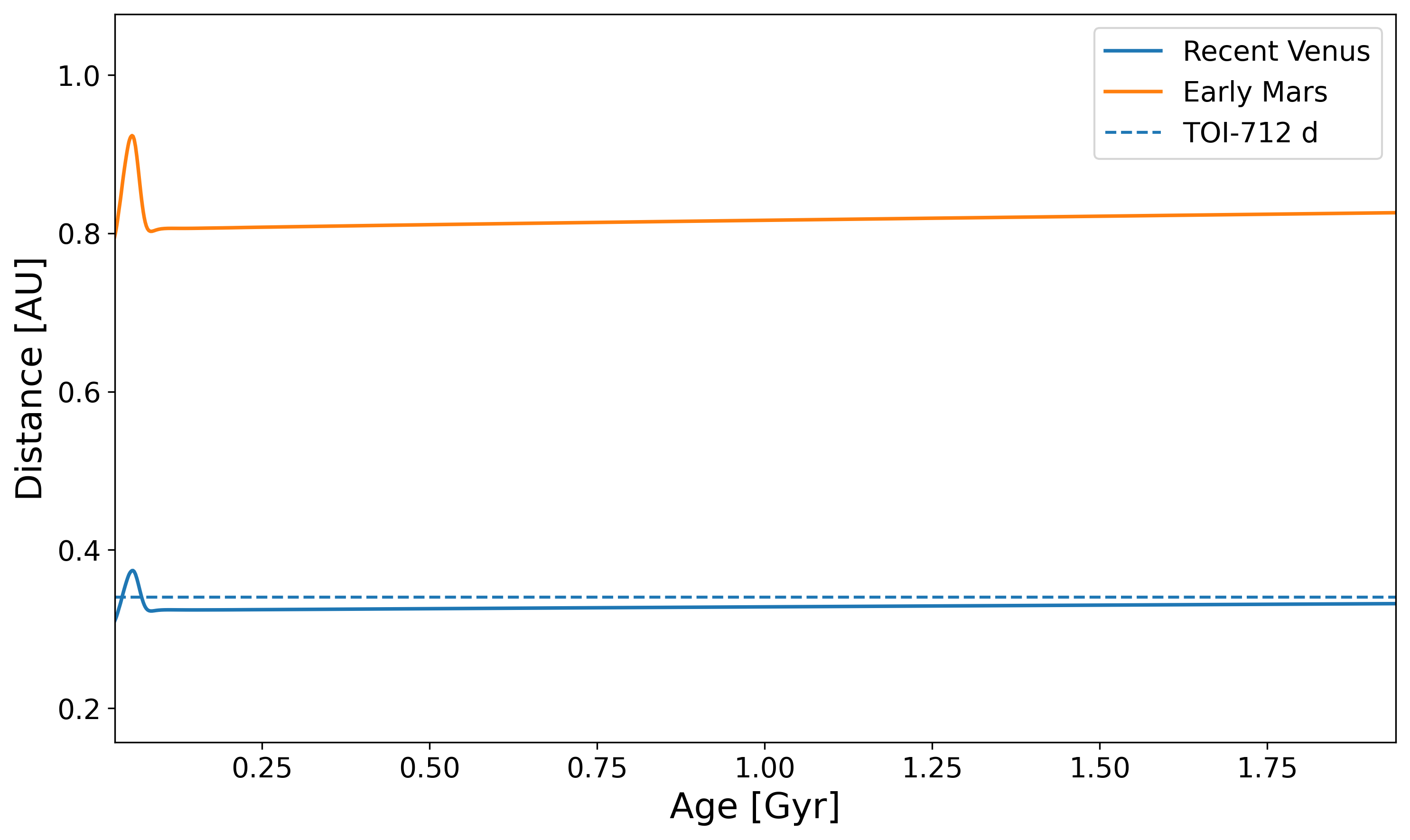}
    \includegraphics[width=\columnwidth]{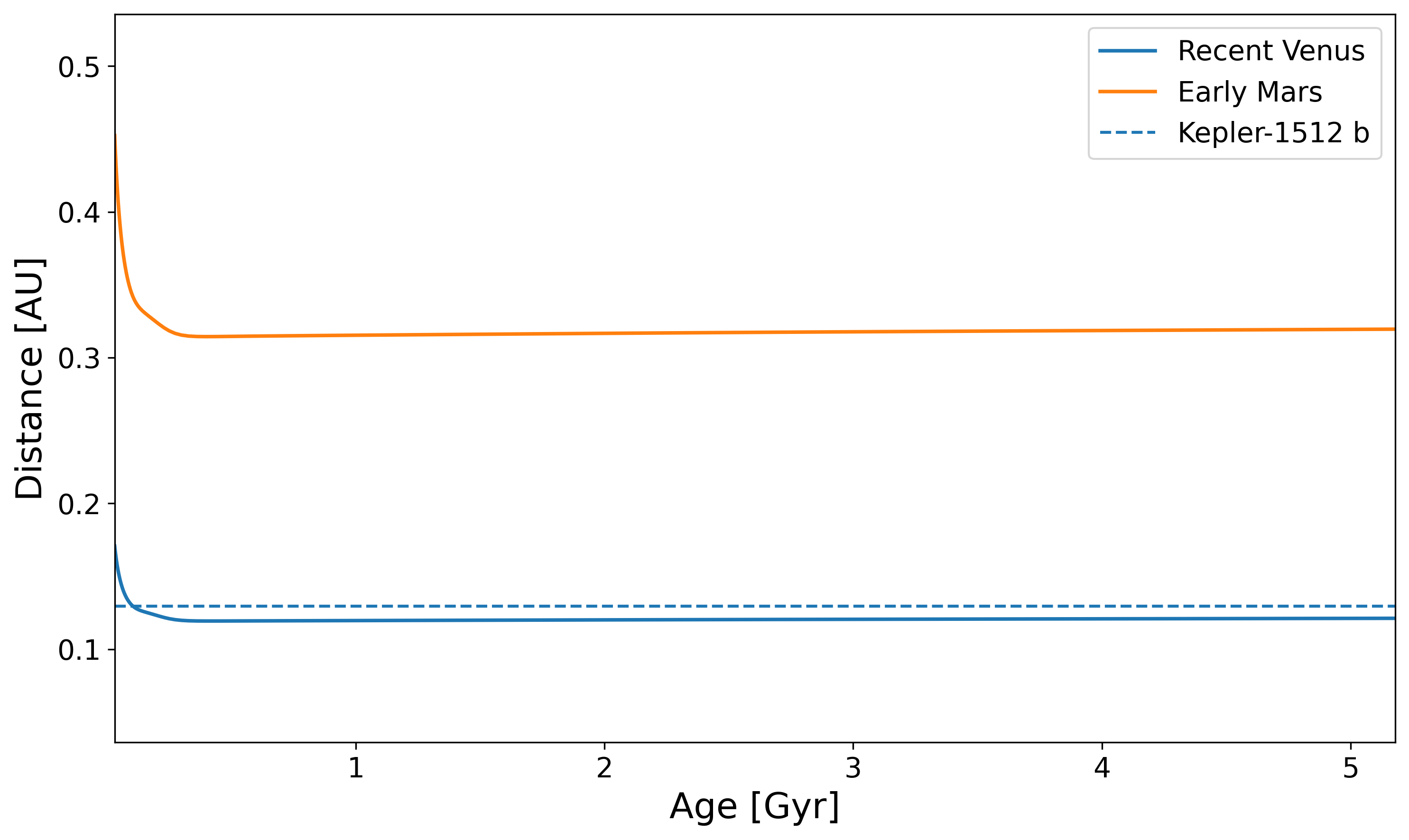}
    \caption{HZ migration histories illustrating the three evolutionary classes defined in this work. Solid curves show the evolution of the Recent Venus (inner, blue line) and Early Mars (outer, orange line) HZ boundaries computed along PARSEC evolutionary sequences while dashed horizontal lines indicate the planets’ current semi-major axes. \textit{Top Left:} HD~27969, a $1.16\,M_\odot$ and $5966\,K$ main-sequence star, hosts a Jovian planet that has continuously resided within the HZ over the interval considered. \textit{Top Right:} HD~132406, mass of $1.13\,M_\odot$ and effective temperature of $5783\,K$, hosts a Jovian planet that entered as the HZ migrated outward. \textit{Bottom Left:} TOI-712, a K-type star with a $0.732\,M_\odot$ mass and $4622\,K$ effective temperature hosts a Neptune-like planet that was in the HZ at earlier times before being desiccated, then moving back between the bounds. \textit{Bottom Right:} Kepler-1512 is a K-type $0.418\,M_\odot$ star with an effective temperature of $4463\,K$ that hosts a super-earth exoplanet. At $30 \, \rm Myr$ the planet, if formed by then, orbited interior to the host's recent Venus boundary which would likely evaporate the surface volatiles.}
    \label{fig:HZRes}
\end{figure*}

Based on this comparison, we separate the planets into three categories. Continuous HZ planets have remained within the HZ throughout the interval considered and therefore represent systems with the longest uninterrupted exposure to temperate stellar irradiation. Cold Start planets formed exterior to the HZ and entered it later as stellar luminosity increased, implying that any surface volatiles would initially have been frozen. Desiccated planets experienced periods interior to the inner HZ boundary, where elevated stellar fluxes could have driven runaway or moist greenhouse conditions and substantial water loss. Figure \ref{fig:HZRes} illustrates representative examples of each class.

Across the full sample, we also quantify the most recent continuous duration each planet has spent within the HZ. Short residence times, particularly those less than $\sim500\,\mathrm{Myr}$, may be insufficient for the emergence of life based on Earth's history (see \cite{Camprub2019} and references therein for an overview of the emergence of life on Earth). This metric, when combined with the HZ migration classification, provides a physically motivated way to prioritize systems for which long-term temperate conditions are most plausible. Table~\ref{tab:HZresidency} summarizes the HZ classification and recent residence time for each system, highlighting how improved stellar ages directly sharpen constraints on planetary climate histories. We find that 97 planets are continuous HZ planets, 37 planets are cold start planets, and 9 planets are likely desiccated. We discuss the implications of these results in Section \ref{sec:Implications_of_HZ}.

\section{Discussion} \label{sec: Discussion}

\begin{figure*}[ht] 
    \centering 
    \begin{minipage}[t]{0.48\textwidth} 
        \centering 
        \includegraphics[width=\linewidth]{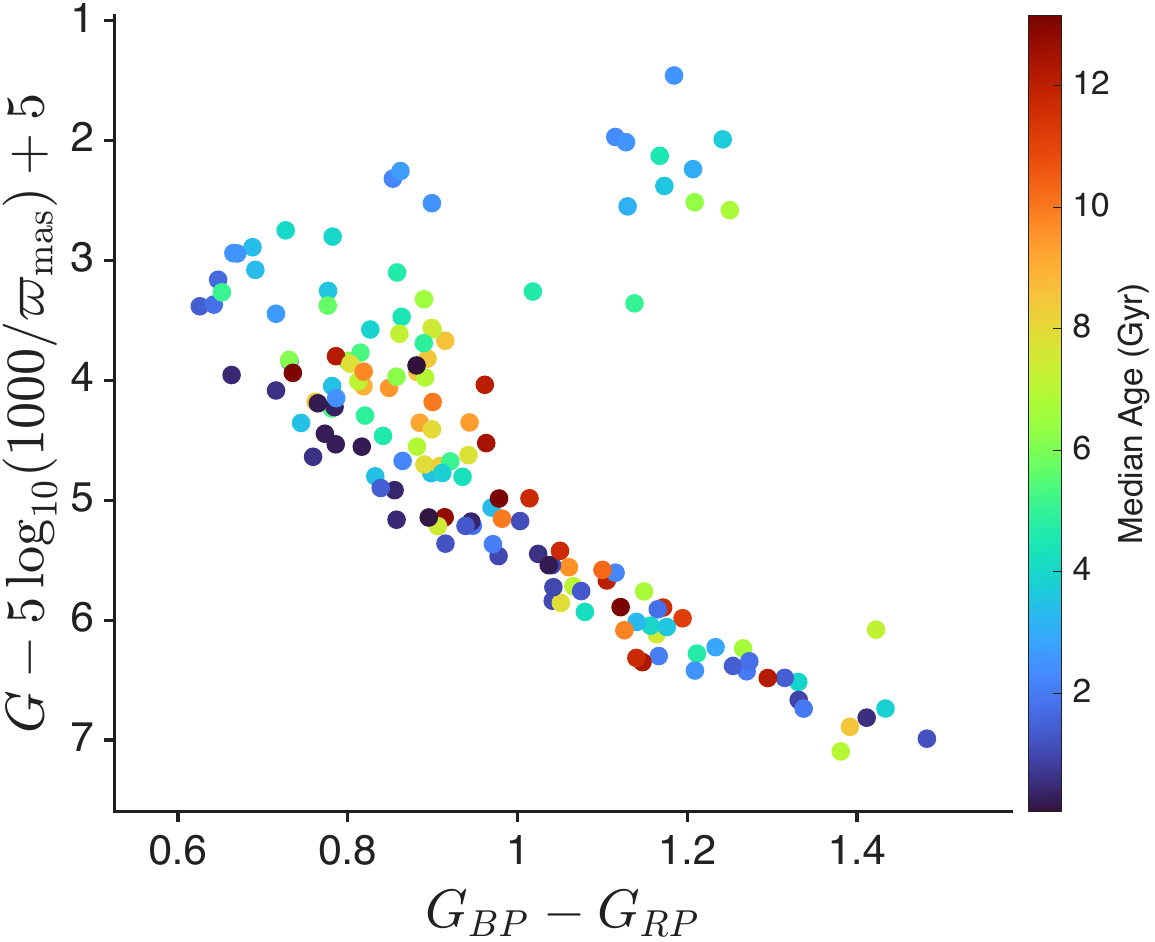}  \caption{Median BASE-9 Age on a Gaia CMD.} \label{fig:MedianAge} 
    \end{minipage} 
    \hfill
    \begin{minipage}[t]{0.48\textwidth} 
        \centering 
        \includegraphics[width=\linewidth]{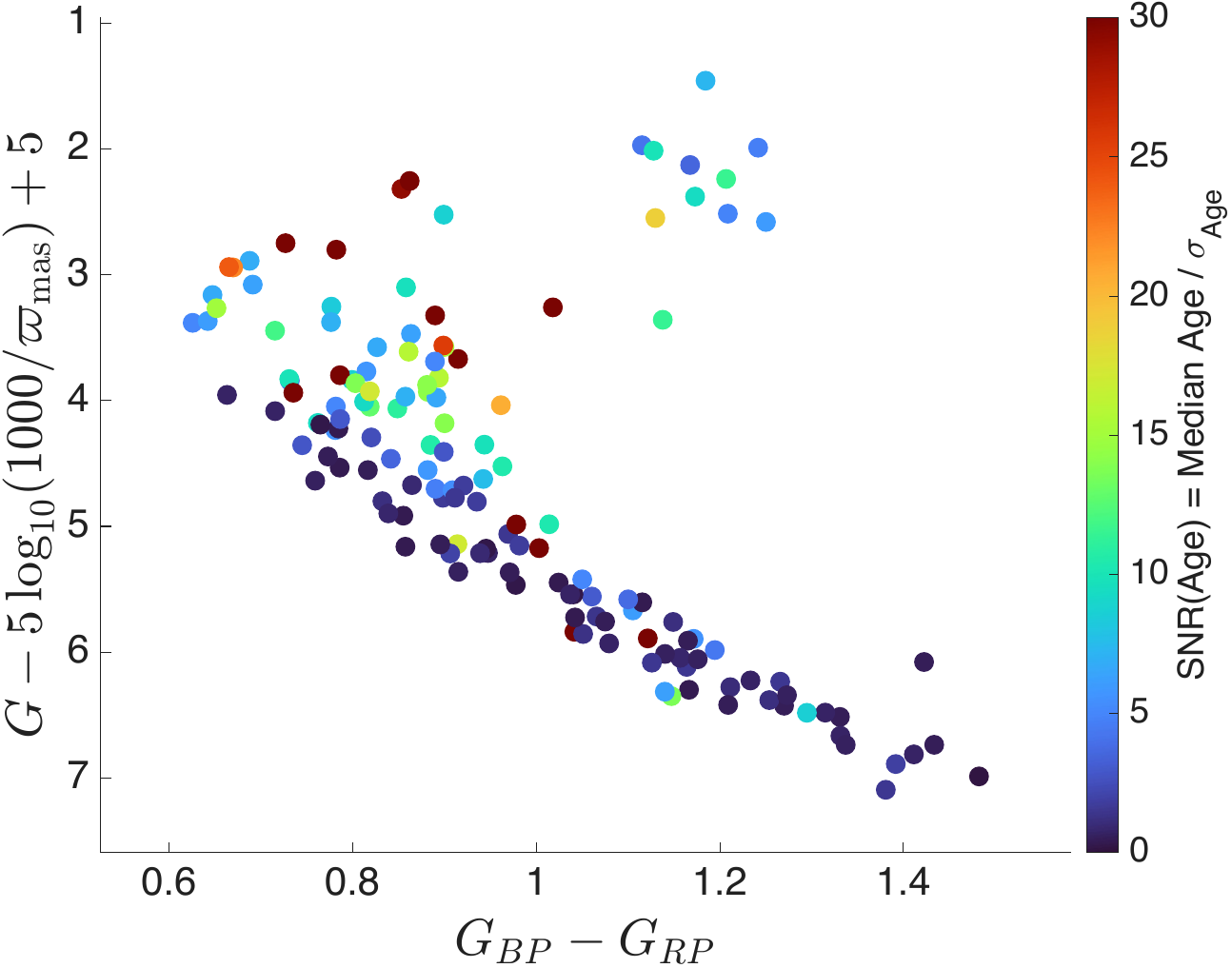} 
        \caption{SNR for the age derivation for each star on a Gaia CMD.} 
        \label{fig:SNR} 
    \end{minipage} 
\end{figure*}

A holistic synthesis of the distribution of stellar median ages and age precision within the Gaia CMD (Figures \ref{fig:CMDs}, \ref{fig:MedianAge}, and \ref{fig:SNR}) clarifies not only where ages are well constrained, but why those constraints emerge. We characterize age precision as SNR(age) = median age divided by $\sigma_{68}$. High SNR(age) stars concentrate along tight loci, and their photometry maps cleanly onto isochrones, yielding narrow, unimodal posteriors under the adopted priors. By contrast, low SNR(age) stars disperse across the CMD, especially on the lower main-sequence where adjacent isochrones converge and the age gradient in CMD space is below the current ability to detect. In this area, uncertainties in metallicity and extinction compete effectively with age, diluting constraining power.

\begin{figure*}[ht]
    \centering
        \includegraphics[width=0.6\textwidth]{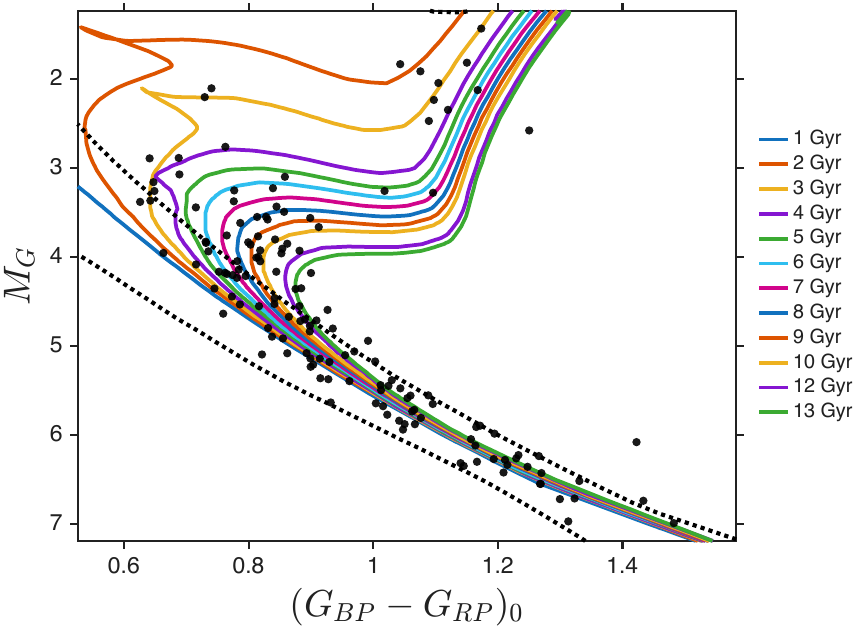}
        \caption{Dereddened Gaia CMD with PARSEC isochrones overplotted. Black points show Gaia DR3 stars corrected for extinction and reddening using band-specific coefficients. Solid curves show isochrones spanning ages from 1 to 13 Gyr at the reference metallicity. The dotted curves highlight 1 Gyr isochrones computed at the lower (-0.4 dex) and upper (+0.4 dex) metallicity values in the sample, illustrating the range of CMD permitted by the observed metallicity distribution. This comparison demonstrates the impact of metallicity on isochrone placement relative to the data.}
    \label{fig:Isochrones}
\end{figure*}

A second complementary pattern appears near the main-sequence turn-off and through the subgiant branch. Even at moderate SNR(age), rapid evolution makes neighboring-age isochrones visibly separate in the CMD as shown in Figure \ref{fig:Isochrones}. However, exactly at the turn-off multiple isochrones and metallicity tracks can overlap or cross, so stars that fall in this intersection often exhibit multimodal posteriors, as highlighted by the concentration of “multimodal” points in Figure \ref{fig:CMDs}. Once a star progresses onto the subgiant branch where isochrones separate, the posteriors typically narrow to a single, well-constrained mode. Practically, ages anchored in the turn-off/subgiant region are robust, while lower main-sequence ages and turn-off intersection cases warrant additional caution or external priors. For exoplanet hosts, this motivates prioritizing evolved systems for follow-up when age-dependent inferences matter most, while treating lower main-sequence ages as lower-fidelity constraints unless supplemented by additional information.

\subsection{Implications of HZ Migration}
\label{sec:Implications_of_HZ}

Placing HZ planets within a time-resolved stellar evolutionary framework adds an important interpretive layer beyond present-day HZ membership. Previous work has shown that stellar luminosity evolution can significantly alter the location of the HZ on Gyr timescales, with consequences for atmospheric retention, surface conditions, and long-term habitability \citep{Kasting1993, Kopparapu2013, Truitt2015}. By anchoring HZ evolution to well-constrained stellar ages and masses, our analysis enables a physically motivated classification of HZ planets based on their residence histories rather than their current orbital configuration alone.

We develop a classification system for planets based on the history of their HZ residency. Planets classified as continuous HZ planets have remained within the HZ throughout the interval considered and are therefore the most likely to preserve primordial surface volatiles. This scenario is broadly consistent with Solar System evolution, where Earth formed during the Sun’s pre–main-sequence phase and maintained clement surface conditions over much of its history, allowing life to emerge within several hundred million years \citep{Nutman2016, Dodd2017}. For these systems, long-term HZ residence strengthens the case for stable climates and sustained habitability.

Cold Start planets, which formed exterior to the HZ and entered it as stellar luminosity increased, present a more ambiguous case. While outward HZ migration can eventually place such planets within classical HZ boundaries, multiple studies suggest that globally glaciated planets may require stellar fluxes exceeding standard HZ limits to deglaciate \citep{Kasting1993, Yang2017}. In some cases, the flux required to thaw a frozen planet could instead trigger rapid water loss via moist greenhouse conditions \citep{Yang2017}. Although narrow windows of stable climates may exist under favorable atmospheric compositions \citep{Wolf2017}, other work argues that limited outgassing may prevent effective deglaciation altogether \citep{Tuchow2020}. As a result, Cold Start planets should not be assumed habitable solely on the basis of present-day HZ membership.

Desiccated planets are those that spent time interior to the inner HZ boundary at any point in their history after planet formation may be complete ($\sim30 \, \rm Myr$). These planets likely have experienced strong stellar irradiation capable of driving substantial water loss. In the prior case, elevated luminosities during early stellar evolution are expected to promote runaway or moist greenhouse states, particularly before planetary atmospheres stabilize \citep{Luger2015, Ramirez2014}. The planet may lose a significant fraction of their primordial volatile inventories and thus require a later-stage delivery of water to be habitable. Kepler-1512 b is an example of this evolutionary history and is shown in Figure \ref{fig:HZRes}.

TOI-712 provides an example of the latter scenario. Because TOI-712 d orbits close to the Recent Venus boundary, the increase in stellar effective temperature as the star stabilizes early in its life produces a brief interval outside the HZ followed by re-entry. This brief period spent interior to the HZ may be sufficient to remove all surface primordial water from the planet.  Figure \ref{fig:HZRes} shows the evolution of the HZ bounds for this system.

Overall, these results reinforce the need to interpret HZ planets in a temporal context. Our framework demonstrates how time-resolved HZ histories, anchored to robust age and mass constraints, can refine target prioritization and sharpen expectations for planetary habitability without overstating definitive outcomes. We suggest that long-lived continuous HZ planets in systems with $\rm SNR(age)>3$ are ideal candidates for future study or even high-resolution follow-ups that are capable of characterizing planetary atmospheres.

\subsection{Literature Age Comparison} \label{sec: litcompare}

\begin{figure*}[t]
    \centering
    \includegraphics[width=0.8\textwidth]{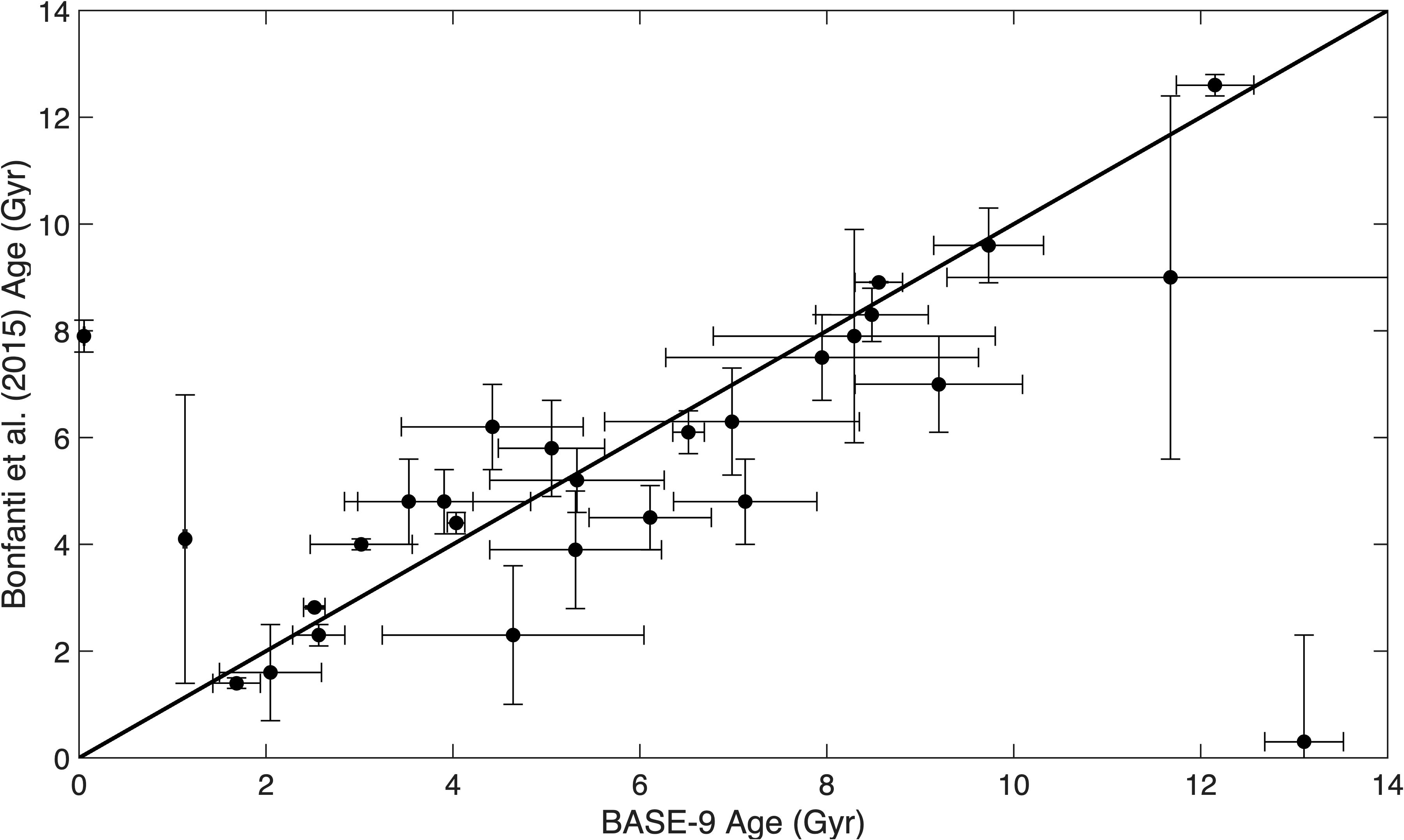}
        \caption{Comparison between ages using BASE-9 and ages from \citet{Bonfanti2015} for the subset of stars with $\mathrm{SNR(age)} > 3$. Error bars show the reported uncertainties in each study. The solid black line indicates the one-to-one relation.}
    \label{fig:litbaseages}
\end{figure*}

We compared our BASE-9-derived stellar ages to the values reported in \cite{Bonfanti2015}'s Table 4. Between our selected HZ host stars and their star's with planets (SWP), there are a common 55 hosts. The comparison between our age values with $SNR(age) > 3$ and theirs is shown in Figure \ref{fig:litbaseages}.

\citet{Bonfanti2015} determined the ages of 317 stars with known planets using PARSEC v1.0 stellar models. \citet{Bonfanti2015} applied two techniques, isochrone placement and Bayesian estimation, but ultimately adopted the isochrone placement results, which assign each star a single best fit position. Their method incorporated safeguards such as filtering on stellar activity and evaluating evolutionary speed in the color–magnitude diagram, but it compresses all uncertainty into a single value rather than retaining the full probability distribution.

Our analysis builds on this foundation using newer data, updated models, and a fully Bayesian approach. BASE-9 uses MCMC sampling to explore the entire posterior probability distribution for each parameter, preserving the correlations among age, metallicity, and extinction. We used PARSECv1.2S isochrones and combined Gaia DR3 parallaxes with Gaia, Pan-STARRS, and 2MASS photometry. Together, these datasets provide significantly higher precision in both parallax and photometric measurements than those available a decade ago, leading to smaller uncertainties in our derived ages. In addition, our analysis does not apply external constraints on stellar mass or $\log g$, allowing these values to be determined self-consistently from the data.

Unlike \citet{Bonfanti2015}'s broader sample of stars with planets of all orbital configurations, our study focuses exclusively on stars that host planets in their HZs. This targeted approach makes our catalog directly relevant to planetary system evolution and long-term habitability. 

Overall, our results agree with the results of \citet{Bonfanti2015} within the uncertainties for the vast majority of stars and more broadly with ages for these systems from the published literature within a few Gyr, although some show systematic differences for certain evolutionary phases. For some main-sequence stars, our Bayesian results trend younger than previous isochrone placement estimates, reflecting broader posteriors that capture the full range of plausible evolutionary tracks. For evolved and turn-off stars, we often find slightly older ages, which likely arise from improvements in stellar models and prior data. The combination of precise photometry, Gaia astrometry, updated PARSEC models, and a Bayesian statistical framework represents a step forward in constraining stellar ages for HZ exoplanet hosts.

\subsection{Limitations}

Although our analysis provides strong constraints on several key stellar properties, there are inherent limitations to our approach. We assumed all stars in our sample to be single-star systems based on their classifications in the NASA Exoplanet Archive \textit{ps} table and the lack of known binarity reported in the literature. Unresolved companions could bias the photometry, making stars appear brighter and redder than assumed and thereby affecting the inferred stellar parameters.

In several cases, we imposed minimum uncertainty floors to obtain well-behaved posterior distributions. Metallicities with reported uncertainties below 0.05 increased to 0.05 and a minimum photometric uncertainty of 0.01 was adopted. While these adjustments are intended to account for unmodeled uncertainties in stellar evolution physics, they introduce an element of subjectivity and underscore that modern, high-precision data can exceed the fidelity of current stellar models.

Despite these mitigations, three stars (Kepler-1058, Kepler-1653, and Kepler-351) consistently failed to yield well-constrained posterior distributions regardless of parameter tuning. All lie along the lower main sequence, where the isochrones for different ages converge, and the photometric evolution over Gyr timescales is minimal. In this regime, stellar ages are fundamentally difficult to constrain using isochrone fitting alone, and the resulting posteriors remain broad and uninformative.

Our HZ migration analysis introduces additional assumptions. We model HZ evolution beginning at $30 \, \rm Myr$, excluding earlier pre–main-sequence phases when rapidly changing stellar luminosities and heightened dynamical activity reduce the likelihood of stable planetary surface environments. Planetary orbits are assumed to be fixed and circular, therefore, excluding the effects of orbital migration, eccentricity-driven flux variations, and dynamical interactions. The HZ boundaries are computed using one-dimensional climate models with fixed atmospheric assumptions; alternative atmospheric compositions or climate feedback could shift these limits. As a result, our HZ classifications provide evolutionary context rather than definitive assessments of habitability and are most robust for evolved and turn-off hosts, where stellar ages are best constrained.

Additional simplifications affect a subset of systems. When stellar parameters such as effective temperature, luminosity, or radius were unavailable in the \textit{ps} table, they were inferred from mass estimates under the assumption of solar density. Orbital distances were calculated assuming zero eccentricity and a solar-mass host where required. 

\section{Conclusions}

In this study, we used BASE-9 to infer posterior distributions for ages and associated stellar parameters for confirmed exoplanet hosts with planets in the habitable zone, combining Gaia, 2MASS, and Pan-STARRS photometry within a Bayesian framework. Our goal was to identify where isochrone-based age estimates are reliable and to quantify the precision achievable across evolutionary stages.

We find that age constraints are strongest near the main-sequence turn-off and along the subgiant branch, where isochrones separate cleanly in color–magnitude space and yield narrow, typically unimodal posteriors. Many stars in these regimes achieve $\geq5\sigma$ age precision, some exceeding $\geq10\sigma$. In contrast, lower main-sequence stars exhibit broader posteriors due to isochrone convergence, resulting in less precise but still informative age estimates. Mapping SNR(age) onto the CMD provides a practical diagnostic for identifying stars with robust ages.

Applying these age constraints in a planetary context allows us to place HZ exoplanets within their host star evolutionary histories. By tracking HZ migration over time, we distinguish planets that have continuously resided within the HZ from those that entered it later or experienced early over-irradiation. This time-resolved perspective adds critical context beyond present-day HZ membership, particularly for evolved and turn-off hosts where stellar ages are most precise.

Overall, our results refine stellar ages for HZ exoplanet hosts and clarify where isochrone fitting is most informative. This work advances stellar characterization while providing a framework for interpreting planetary habitability in an evolutionary context, supporting efforts to prioritize targets in the search for life beyond the Solar System.

\section{Acknowledgments}

This work was supported by funding from the Office of Undergraduate Research and the Dean's Office of the College of Arts and Sciences at Embry-Riddle Aeronautical University. The authors also gratefully acknowledge travel support provided by the Department of Physical Sciences. The authors thank Elliot Robinson, Elizabeth Jeffery, David Stenning, and David van Dyk for their valuable input, guidance, and expertise throughout the course of this project.

\clearpage
\appendix 

\input{BASE-9Results.txt}

\input{HZ_Residency.txt}

\input{PriorTable.txt}

\clearpage

\bibliographystyle{aasjournal}
\bibliography{MainDocument}

\end{document}

%% file: BASE-9Results.txt
\begin{longtable}{lcccccc}
\caption{BASE-9 Results} 
\label{tab:results} \\

\toprule
Host Star & Age (Gyr) & [Fe/H] & Parallax (mas) & $A_{g}$ (mmag) & Mass ($M_{\odot}$) & ID Code \\
\midrule
\endfirsthead

\toprule
Host Star & Age (Gyr) & [Fe/H] & Parallax (mas) & $A_{g}$ (mmag) & Mass ($M_{\odot}$) & ID Code \\
\midrule
\endhead

\midrule
\multicolumn{7}{r}{{Continued on next page}} \\
\endfoot

\bottomrule

\endlastfoot

BD+14 4559 & $11.7^{+1.27}_{-2.39}$ & $0.268^{+0.036}_{-0.0373}$ & $20.3^{+0.014}_{-0.015}$ & $0.323^{+0.291}_{-0.207}$ & $0.895^{+0.00651}_{-0.0177}$ & 41112 \\ 
BD+55 362 & $9.62^{+2.49}_{-3.4}$ & $0.196^{+0.0444}_{-0.0425}$ & $19^{+0.014}_{-0.014}$ & $1.67^{+1.03}_{-0.89}$ & $0.899^{+0.036}_{-0.0121}$ & 41111 \\ 
BD-08 2823 & $1.89^{+5.05}_{-1.72}$ & $0.104^{+0.0741}_{-0.0712}$ & $24.2^{+0.015}_{-0.013}$ & $0.241^{+0.255}_{-0.168}$ & $0.781^{+0.0153}_{-0.015}$ & 31111 \\ 
HAT-P-13 & $4.42^{+0.973}_{-0.424}$ & $0.9^{+0.0477}_{-0.0351}$ & $4.08^{+0.018}_{-0.019}$ & $34.9^{+4.18}_{-4.11}$ & $1.28^{+0.0208}_{-0.0786}$ & 21111 \\ 
HD 100777 & $3.44^{+1.85}_{-2.32}$ & $0.286^{+0.0537}_{-0.0507}$ & $20.2^{+0.027}_{-0.03}$ & $0.243^{+0.223}_{-0.167}$ & $1.04^{+0.0337}_{-0.0267}$ & 11111 \\ 
HD 10180 & $5.31^{+0.903}_{-0.919}$ & $0.0831^{+0.0496}_{-0.0493}$ & $25.7^{+0.015}_{-0.015}$ & $0.069^{+0.072}_{-0.048}$ & $1.07^{+0.0178}_{-0.0126}$ & 11111 \\ 
HD 103891 & $4.06^{+0.127}_{-0.104}$ & $-0.166^{+0.0445}_{-0.0424}$ & $18.2^{+0.039}_{-0.039}$ & $0^{+0}_{-0}$ & $1.24^{+0.0158}_{-0.0144}$ & 11111 \\
HD 103949 & $4.49^{+5.43}_{-3.3}$ & $-0.00915^{+0.0411}_{-0.0459}$ & $37.7^{+0.029}_{-0.03}$ & $0.034^{+0.036}_{-0.024}$ & $0.78^{+0.012}_{-0.0134}$ & 51111 \\ 
HD 10442 & $2.39^{+0.607}_{-0.5}$ & $-0.0245^{+0.0511}_{-0.0487}$ & $7.51^{+0.028}_{-0.028}$ & $79.6^{+85.6}_{-56.5}$ & $1.56^{+0.13}_{-0.121}$ & 11111 \\ 
HD 106270 & $3.02^{+0.0265}_{-0.546}$ & $0.142^{+0.07}_{-0.0764}$ & $10.5^{+0.033}_{-0.031}$ & $316^{+7.59}_{-6.56}$ & $1.48^{+0.108}_{-0.0253}$ & 21111 \\ 
HD 108874 & $7.95^{+1.52}_{-1.67}$ & $0.201^{+0.05}_{-0.05}$ & $16.8^{+0.242}_{-0.252}$ & $0.135^{+0.141}_{-0.092}$ & $1.02^{+0.0688}_{-0.0155}$ & 11111 \\ 
HD 109286 & $0.428^{+1.6}_{-0.359}$ & $0.0361^{+0.0368}_{-0.0417}$ & $18.1^{+0.025}_{-0.024}$ & $2.47^{+2.6}_{-1.72}$ & $1.01^{+0.0145}_{-0.0278}$ & 31111 \\ 
HD 111998 & $1.47^{+0.288}_{-0.3}$ & $0.1^{+0.0446}_{-0.0484}$ & $29.9^{+0.037}_{-0.037}$ & $0.034^{+0.037}_{-0.023}$ & $1.31^{+0.0315}_{-0.0191}$ & 11111 \\ 
HD 114783 & $7^{+3.75}_{-5.84}$ & $0.0565^{+0.0471}_{-0.0454}$ & $47.6^{+0.027}_{-0.03}$ & $0.275^{+0.284}_{-0.193}$ & $0.858^{+0.0253}_{-0.0227}$ & 51111 \\ 
HD 11506 & $2.05^{+0.492}_{-0.546}$ & $0.345^{+0.0434}_{-0.0513}$ & $19.5^{+0.022}_{-0.022}$ & $0.034^{+0.037}_{-0.024}$ & $1.23^{+0.0227}_{-0.0166}$ & 11112 \\ 
HD 124330 & $6.39^{+0.577}_{-0.829}$ & $0.228^{+0.0564}_{-0.0529}$ & $16.6^{+0.014}_{-0.015}$ & $0.118^{+0.089}_{-0.07}$ & $1.16^{+0.0154}_{-0.0223}$ & 11111 \\ 
HD 128311 & $7.32^{+3.97}_{-5.42}$ & $0.00371^{+0.0396}_{-0.0398}$ & $61.3^{+0.042}_{-0.042}$ & $0.034^{+0.037}_{-0.024}$ & $0.8^{+0.0142}_{-0.0192}$ & 51111 \\ 
HD 128356 & $1.71^{+5.2}_{-1.57}$ & $0.286^{+0.052}_{-0.0513}$ & $38.4^{+0.028}_{-0.028}$ & $0.213^{+0.205}_{-0.145}$ & $0.874^{+0.0127}_{-0.0195}$ & 31112 \\ 
HD 13167 & $2.27^{+0.0923}_{-0.0629}$ & $0.278^{+0.0358}_{-0.0355}$ & $6.75^{+0.023}_{-0.02}$ & $211^{+3.86}_{-3.84}$ & $1.65^{+0.016}_{-0.0173}$ & 11112 \\ 
HD 132406 & $8.99^{+0.682}_{-0.688}$ & $0.111^{+0.0497}_{-0.0495}$ & $14.2^{+0.019}_{-0.02}$ & $0.323^{+0.294}_{-0.213}$ & $1.13^{+0.00665}_{-0.0201}$ & 11111 \\ 
HD 134606 & $6.97^{+1.18}_{-1.12}$ & $0.302^{+0.049}_{-0.0504}$ & $37.3^{+0.018}_{-0.018}$ & $0.604^{+0.464}_{-0.372}$ & $1.04^{+0.013}_{-0.00684}$ & 11111 \\ 
HD 13724 & $0.23^{+0.712}_{-0.168}$ & $0.22^{+0.0348}_{-0.0362}$ & $23^{+0.017}_{-0.018}$ & $0.033^{+0.037}_{-0.023}$ & $1.11^{+0.0129}_{-0.0196}$ & 31112 \\ 
HD 137388 & $0.668^{+2.53}_{-0.572}$ & $0.219^{+0.0408}_{-0.0441}$ & $24.6^{+0.013}_{-0.013}$ & $0.033^{+0.037}_{-0.023}$ & $0.942^{+0.0138}_{-0.0251}$ & 31111 \\ 
HD 137496 & $8.44^{+0.511}_{-0.524}$ & $-0.0264^{+0.0489}_{-0.0501}$ & $6.42^{+0.019}_{-0.018}$ & $202^{+26.8}_{-27.6}$ & $1.2^{+0.0209}_{-0.021}$ & 11112 \\ 
HD 13908 & $3.86^{+0.308}_{-0.689}$ & $-0.00489^{+0.0843}_{-0.0487}$ & $12.5^{+0.021}_{-0.02}$ & $0.293^{+0.263}_{-0.184}$ & $1.3^{+0.0601}_{-0.0594}$ & 21111 \\ 
HD 141399 & $9.95^{+0.678}_{-0.695}$ & $0.302^{+0.0497}_{-0.0475}$ & $27^{+0.015}_{-0.015}$ & $0.503^{+0.488}_{-0.342}$ & $1.13^{+0.0143}_{-0.0129}$ & 11111 \\ 
HD 141937 & $0.295^{+1.13}_{-0.239}$ & $0.142^{+0.0334}_{-0.0389}$ & $30.6^{+0.087}_{-0.088}$ & $0.068^{+0.074}_{-0.048}$ & $1.1^{+0.0143}_{-0.0289}$ & 31111 \\ 
HD 142415 & $0.277^{+0.769}_{-0.218}$ & $0.137^{+0.0343}_{-0.0381}$ & $28.2^{+0.01}_{-0.011}$ & $0.068^{+0.073}_{-0.048}$ & $1.12^{+0.0133}_{-0.0246}$ & 31111 \\ 
HD 145934 & $2.61^{+0.384}_{-0.312}$ & $0.0731^{+0.0603}_{-0.0615}$ & $4.37^{+0.022}_{-0.023}$ & $20.8^{+2.18}_{-2.26}$ & $1.55^{+0.0797}_{-0.0825}$ & 11112 \\ 
HD 153950 & $6.11^{+0.562}_{-0.654}$ & $-0.00486^{+0.05}_{-0.05}$ & $20.7^{+0.024}_{-0.025}$ & $0.067^{+0.075}_{-0.047}$ & $1.12^{+0.0154}_{-0.00992}$ & 11112 \\  
HD 155193 & $3.36^{+0.792}_{-0.247}$ & $0.042^{+0.0576}_{-0.0462}$ & $17.3^{+0.02}_{-0.02}$ & $6.09^{+5.03}_{-3.97}$ & $1.3^{+0.0297}_{-0.0676}$ & 21111 \\ 
HD 155358 & $13.1^{+0.246}_{-0.421}$ & $-0.617^{0.0424}_{-0.0392}$ & $22.9^{+0.014}_{-0.013}$ & $0.329^{+0.301}_{-0.205}$ & $1.17^{+0.00643}_{-0.0579}$ & 41112 \\
HD 156411 & $4.03^{+0.0934}_{-0.0777}$ & $-0.0982^{+0.0468}_{-0.0499}$ & $18^{+0.022}_{-0.019}$ & $36.5^{+3}_{-2.87}$ & $1.26^{+0.0146}_{-0.0259}$ & 11111 \\ 
HD 159868 & $6.52^{+0.171}_{-0.171}$ & $-0.0456^{0.0526}_{-0.0521}$ & $17.9^{+0.022}_{-0.021}$ & $93.1^{+14.0}_{-14.7}$ & $1.34^{+0.00397}_{-0.105}$ & 11112 \\
HD 1605 & $3.14^{+0.176}_{-0.158}$ & $0.108^{+0.052}_{-0.0544}$ & $11.4^{+0.02}_{-0.021}$ & $75.4^{+3.23}_{-3.4}$ & $1.46^{+0.0398}_{-0.0439}$ & 11112 \\ 
HD 16175 & $3.62^{+0.606}_{-0.171}$ & $0.294^{+0.0547}_{-0.0521}$ & $16.7^{+0.027}_{-0.03}$ & $0.143^{+0.12}_{-0.094}$ & $1.33^{+0.0104}_{-0.041}$ & 11111 \\ 
HD 163607 & $8.56^{+0.258}_{-0.251}$ & $0.221^{+0.0484}_{-0.0578}$ & $14.8^{+0.017}_{-0.018}$ & $12.7^{+12.6}_{-8.17}$ & $1.26^{+0.0202}_{-0.00592}$ & 11112 \\ 
HD 165155 & $1.78^{+2.98}_{-1.66}$ & $0.134^{+0.0436}_{-0.0488}$ & $15.8^{+0.02}_{-0.022}$ & $33.2^{+3.72}_{-3.65}$ & $0.962^{+0.0268}_{-0.0283}$ & 31111 \\ 
HD 169830 & $2.52^{+0.111}_{-0.115}$ & $0.196^{+0.0505}_{-0.0467}$ & $27.2^{+0.146}_{-0.147}$ & $0^{+0}_{-0}$ & $1.4^{+0.015}_{-0.0145}$ & 11112 \\ 
HD 175167 & $6.72^{+0.429}_{-0.489}$ & $0.336^{+0.0767}_{-0.0702}$ & $14^{+0.024}_{-0.022}$ & $80.9^{+15.4}_{-15.8}$ & $1.28^{+0.0144}_{-0.0831}$ & 11112 \\ 
HD 17674 & $8.37^{+0.887}_{-0.874}$ & $-0.167^{+0.0469}_{-0.0468}$ & $22.5^{+0.026}_{-0.025}$ & $0.034^{+0.036}_{-0.024}$ & $1.08^{+0.00262}_{-0.0167}$ & 11111 \\ 
HD 18015 & $2.76^{+0.0259}_{-0.0267}$ & $-0.0912^{+0.0494}_{-0.0493}$ & $8.01^{+0.023}_{-0.021}$ & $0^{+0}_{-0}$ & $1.46^{+0.0131}_{-0.0146}$ & 11111 \\ 
HD 181720 & $12.2^{+0.39}_{-0.414}$ & $-0.513^{+0.0469}_{-0.0497}$ & $16.7^{+0.026}_{-0.027}$ & $40^{+5.48}_{-5.65}$ & $1.19^{+0.0202}_{-0.0153}$ & 11112 \\ 
HD 183263 & $3.53^{+0.687}_{-0.68}$ & $0.309^{+0.0514}_{-0.0504}$ & $18.3^{+0.02}_{-0.021}$ & $0.148^{+0.128}_{-0.097}$ & $1.15^{+0.0184}_{-0.0209}$ & 11112 \\ 
HD 190647 & $8.48^{+0.576}_{-0.602}$ & $0.257^{+0.0494}_{-0.0494}$ & $18.4^{+0.024}_{-0.024}$ & $1.98^{+2.02}_{-1.36}$ & $1.18^{+0.00847}_{-0.00957}$ & 11111 \\ 
HD 191939 & $12.7^{+0.534}_{-0.903}$ & $-0.13^{+0.0321}_{-0.0302}$ & $18.7^{+0.014}_{-0.013}$ & $0.147^{+0.133}_{-0.098}$ & $0.911^{+0.00576}_{-0.0112}$ & 42111 \\ 
HD 2039 & $0.308^{+1.36}_{-0.275}$ & $0.335^{+0.0481}_{-0.0453}$ & $11.8^{+0.013}_{-0.014}$ & $19.3^{+4.76}_{-4.87}$ & $1.2^{+0.0288}_{-0.0364}$ & 31111 \\ 
HD 208487 & $0.633^{+1.11}_{-0.595}$ & $0.118^{+0.0441}_{-0.0493}$ & $22.3^{+0.029}_{-0.031}$ & $0.066^{+0.077}_{-0.047}$ & $1.18^{+0.0275}_{-0.0384}$ & 31112 \\ 
HD 20868 & $2.82^{+5.28}_{-2.4}$ & $0.113^{+0.0444}_{-0.0432}$ & $21^{+0.013}_{-0.013}$ & $1.39^{+1.14}_{-0.906}$ & $0.805^{+0.0153}_{-0.0126}$ & 51112 \\ 
HD 210277 & $8.29^{+1.46}_{-1.51}$ & $0.196^{+0.0484}_{-0.0496}$ & $46.9^{+0.028}_{-0.028}$ & $0.033^{+0.037}_{-0.023}$ & $1.02^{+0.0681}_{-0.0132}$ & 11111 \\ 
HD 216437 & $5.33^{+0.934}_{-0.889}$ & $0.267^{+0.0373}_{-0.0471}$ & $37.5^{+0.025}_{-0.024}$ & $0.102^{+0.109}_{-0.072}$ & $1.18^{+0.0172}_{-0.0272}$ & 11111 \\
HD 216520 & $1.08^{+0.029}_{-0.0276}$ & $-0.312^{+0.0337}_{-0.0322}$ & $51.2^{+0.02}_{-0.018}$ & $0^{+0}_{-0}$ & $0.842^{+0.00525}_{-0.00482}$ & 11111 \\ 
HD 218566 & $11.3^{+1.61}_{-2.93}$ & $0.277^{+0.0412}_{-0.0409}$ & $34.7^{+0.029}_{-0.028}$ & $0.034^{+0.037}_{-0.024}$ & $0.878^{+0.0074}_{-0.03}$ & 41112 \\ 
HD 219415 & $6.32^{+1.42}_{-1.12}$ & $0.00394^{+0.0484}_{-0.0491}$ & $6.04^{+0.014}_{-0.013}$ & $166^{+1.00}_{-0.983}$ & $1.26^{+0.032}_{-0.0243}$ &  11112 \\ 
HD 221287 & $0.482^{+0.853}_{-0.444}$ & $0.0867^{+0.0429}_{-0.0495}$ & $17.9^{+0.02}_{-0.021}$ & $0.068^{+0.072}_{-0.047}$ & $1.21^{+0.0251}_{-0.0341}$ & 31111 \\ 
HD 221585 & $7.52^{+0.221}_{-0.377}$ & $0.291^{+0.0544}_{-0.0549}$ & $17.924^{+0.016}_{-0.015}$ & $0.063^{+0.071}_{-0.043}$ & $1.29^{+0.00129}_{-0.00369}$ & 11111 \\ 
HD 224538 & $2.68^{+0.228}_{-0.227}$ & $0.368^{+0.0494}_{-0.0473}$ & $12.6^{+0.02}_{-0.02}$ & $0.505^{+0.398}_{-0.319}$ & $1.3^{+0.00967}_{-0.00915}$ & 11111 \\ 
HD 23079 & $3.52^{+1.01}_{-1.48}$ & $-0.0971^{+0.047}_{-0.0498}$ & $29.9^{+0.019}_{-0.02}$ & $0.034^{+0.037}_{-0.023}$ & $1.04^{+0.0401}_{-0.0209}$ & 11111 \\ 
HD 23127 & $3.91^{+0.925}_{-0.235}$ & $0.375^{+0.0547}_{-0.0542}$ & $10.7^{+0.014}_{-0.013}$ & $24.3^{+1.61}_{-1.6}$ & $1.26^{+0.00899}_{-0.0407}$ & 21111 \\ 
HD 24040 & $7.13^{+0.743}_{-0.768}$ & $0.223^{+0.0521}_{-0.0488}$ & $21.5^{+0.024}_{-0.024}$ & $0.425^{+0.395}_{-0.286}$ & $1.13^{+0.0113}_{-0.0113}$ & 11111 \\ 
HD 27969 & $5.7^{+0.385}_{-1.24}$ & $0.188^{+0.0567}_{-0.056}$ & $14.8^{+0.028}_{-0.027}$ & $1.02^{+0.952}_{-0.698}$ & $1.26^{+0.0272}_{-0.0726}$ & 21111 \\ 
HD 28185 & $2.28^{+2.16}_{-2.23}$ & $0.229^{+0.0506}_{-0.0507}$ & $25.5^{+0.019}_{-0.021}$ & $0.567^{+0.549}_{-0.372}$ & $1.06^{+0.0375}_{-0.0362}$ & 31111 \\ 
HD 33564 & $1.82^{+0.269}_{-0.318}$ & $0.259^{+0.0492}_{-0.0523}$ & $48.1^{+0.07}_{-0.071}$ & $0.033^{+0.037}_{-0.023}$ & $1.33^{+0.0189}_{-0.017}$ & 11111 \\ 
HD 34445 & $7.81^{+0.556}_{-0.568}$ & $0.152^{+0.0492}_{-0.049}$ & $21.8^{+0.02}_{-0.021}$ & $0.288^{+0.261}_{-0.188}$ & $1.22^{+0.0143}_{-0.066}$ & 11112 \\ 
HD 38801 & $4.7^{+0.107}_{-0.0662}$ & $0.245^{+0.047}_{-0.0492}$ & $11^{+0.021}_{-0.022}$ & $0^{+0}_{-0}$ & $1.28^{+0.0128}_{-0.0173}$ & 11111 \\  
HD 40307 & $2.15^{+8.88}_{-2.101}$ & $-0.298^{+0.0473}_{-0.0516}$ & $77.3^{+0.018}_{-0.017}$ & $0^{+0}_{-0}$ & $0.710^{+0.0599}_{-0.0138}$ & 51111 \\
HD 4203 & $6.99^{+0.68}_{-1.36}$ & $0.387^{+0.0627}_{-0.0533}$ & $12.3^{+0.02}_{-0.019}$ & $69.4^{+2.61}_{-2.6}$ & $1.17^{+0.00999}_{-0.0152}$ & 11111 \\
HD 43197 & $4.22^{+1.77}_{-2.94}$ & $0.411^{+0.0487}_{-0.0503}$ & $16^{+0.011}_{-0.01}$ & $0.871^{+0.802}_{-0.569}$ & $1.03^{+0.0408}_{-0.0216}$ & 11111 \\
HD 44219 & $9.73^{+0.535}_{-0.587}$ & $0.0497^{+0.0477}_{-0.0472}$ & $18.9^{+0.018}_{-0.019}$ & $0.589^{+0.453}_{-0.371}$ & $1.21^{+0.00611}_{-0.0373}$ & 11111 \\ 
HD 45350 & $9.2^{+0.897}_{-0.837}$ & $0.277^{+0.0491}_{-0.0482}$ & $21.3^{+0.028}_{-0.028}$ & $1.51^{+1.39}_{-0.986}$ & $1.08^{+0.00307}_{-0.00386}$ & 11111 \\ 
HD 45364 & $7.6^{+2.72}_{-3.7}$ & $-0.136^{+0.0544}_{-0.0474}$ & $29.1^{+0.018}_{-0.017}$ & $5.84^{+3.86}_{-3.39}$ & $0.914^{+0.0351}_{-0.0214}$ & 11111 \\ 
HD 48265 & $4.69^{+0.0493}_{-0.819}$ & $0.39^{+0.0566}_{-0.049}$ & $11^{+0.016}_{-0.017}$ & $0^{+0}_{-0}$ & $1.28^{+0.0935}_{-0.00542}$ & 21111 \\ 
HD 48948 & $1.18^{+10.7}_{-0.00645}$ & $-0.0395^{+0.0294}_{-0.0244}$ & $59.4^{+0.025}_{-0.026}$ & $0^{+0}_{-0}$ & $0.725^{+0.00535}_{-0.00527}$ & 22112 \\ 
HD 564 & $0.585^{+1.46}_{-0.512}$ & $-0.158^{+0.0396}_{-0.0515}$ & $19.5^{+0.018}_{-0.018}$ & $0.276^{+0.269}_{-0.188}$ & $1.02^{+0.017}_{-0.0346}$ & 31111 \\ 
HD 63765 & $1.22^{+2.79}_{-1.05}$ & $-0.128^{+0.0413}_{-0.044}$ & $30.8^{+0.017}_{-0.018}$ & $0.178^{+0.164}_{-0.118}$ & $0.897^{+0.0159}_{-0.0245}$ & 31111 \\ 
HD 7199 & $3.38^{+1.98}_{-2.29}$ & $0.348^{+0.0497}_{-0.0518}$ & $27.7^{+0.017}_{-0.015}$ & $0.174^{+0.181}_{-0.125}$ & $0.993^{+0.0266}_{-0.0213}$ & 11112 \\ 
HD 73526 & $0.0536^{+0.00567}_{-0.00192}$ & $-0.499^{+0.036}_{-0.0129}$ & $27.2^{+0.026}_{-0.023}$ & $9.11^{+10.9}_{-6.99}$ & $0.751^{+0.00337}_{-0.00378}$ & 12122 \\ 
HD 73534 & $5.06^{+0.568}_{-0.301}$ & $0.176^{+0.0463}_{-0.0507}$ & $12^{+0.027}_{-0.024}$ & $77.5^{+21.8}_{-20.3}$ & $1.26^{+0.024}_{-0.0227}$ & 11111 \\ 
HD 80653 & $2.48^{+0.815}_{-0.852}$ & $0.342^{+0.0447}_{-0.0502}$ & $9.26^{+0.022}_{-0.022}$ & $5.48^{+4.03}_{-3.38}$ & $1.16^{+0.0275}_{-0.0219}$ & 11111 \\ 
HD 82943 & $0.282^{+1.29}_{-0.249}$ & $0.281^{+0.0444}_{-0.0477}$ & $36.1^{+0.027}_{-0.026}$ & $0.034^{+0.037}_{-0.024}$ & $1.2^{+0.0229}_{-0.0371}$ & 31111 \\ 
HD 8326 & $3.99^{+5.29}_{-3.4}$ & $0.0691^{+0.0431}_{-0.0423}$ & $32.6^{+0.021}_{-0.022}$ & $0.033^{+0.037}_{-0.022}$ & $0.82^{+0.0162}_{-0.0162}$ & 51111 \\ 
HD 86264 & $1.69^{+0.247}_{-0.254}$ & $0.419^{+0.0674}_{-0.0503}$ & $14.9^{+0.022}_{-0.022}$ & $0.424^{+0.378}_{-0.296}$ & $1.408^{+0.0164}_{-0.0149}$ & 11112 \\ 
HD 9174 & $4.93^{+1.31}_{-0.596}$ & $0.408^{+0.0326}_{-0.0435}$ & $12.3^{+0.022}_{-0.022}$ & $110^{+1.84}_{-1.83}$ & $1.24^{+0.0132}_{-0.0223}$ & 21112 \\ 
HD 92788 & $4.64^{+1.16}_{-1.4}$ & $0.287^{+0.0526}_{-0.0496}$ & $28.9^{+0.025}_{-0.027}$ & $0.034^{+0.036}_{-0.024}$ & $1.07^{+0.0275}_{-0.0178}$ & 11111 \\ 
HD 94834 & $3.58^{+0.462}_{-0.289}$ & $0.0106^{+0.0472}_{-0.0434}$ & $10.3^{+0.026}_{-0.026}$ & $141^{+2.89}_{-2.97}$ & $1.37^{+0.0492}_{-0.054}$ & 11112 \\ 
HD 95089 & $2.56^{+0.28}_{-0.239}$ & $-0.00237^{+0.0515}_{-0.0488}$ & $7.38^{+0.025}_{-0.024}$ & $96.9^{+4.73}_{-4.77}$ & $1.53^{+0.0617}_{-0.0634}$ & 11111 \\ 
HD 99109 & $1.13^{+0.0185}_{-0.0129}$ & $0.0777^{+0.0479}_{-0.0418}$ & $18.2^{+0.016}_{-0.017}$ & $0.619^{+0.637}_{-0.41}$ & $1.01^{+0.0044}_{-0.00472}$ & 11111 \\ 
HIP 5158 & $6.56^{+4.13}_{-4.44}$ & $0.285^{+0.0555}_{-0.055}$ & $19.3^{+0.02}_{-0.02}$ & $0.033^{+0.037}_{-0.024}$ & $0.822^{+0.0149}_{-0.011}$ & 51111 \\ 
HIP 56640 & $3.04^{+0.26}_{-0.267}$ & $0.0971^{+0.0515}_{-0.0531}$ & $8.22^{+0.019}_{-0.02}$ & $190^{+2.29}_{-2.23}$ & $1.47^{+0.058}_{-0.0575}$ & 11112 \\ 
HIP 57274 & $12.1^{+1.01}_{-0.9.56}$ & $0.0801^{+0.0476}_{-0.0311}$ & $38.7^{+0.024}_{-0.023}$ & $0.182^{+0.167}_{-0.123}$ & $0.773^{+0.00419}_{-0.0047}$ & 21112 \\ 
KELT-6 & $5.07^{+0.311}_{-0.363}$ & $-0.227^{+0.0578}_{-0.0502}$ & $4.11^{+0.017}_{-0.017}$ & $11.2^{+9.93}_{-7.30}$ & $1.21^{+0.0158}_{-0.0534}$ & 11111 \\ 
kap CrB & $4.52^{+1.55}_{-0.851}$ & $0.0725^{+0.0656}_{-0.0671}$ & $33.3^{+0.078}_{-0.079}$ & $0.149^{+0.13}_{-0.097}$ & $1.33^{+0.0761}_{-0.0589}$ & 11112 \\ 
Kepler-1038 & $3.54^{+7.24}_{-3.49}$ & $-0.0897^{+0.0763}_{-0.0910}$ & $1.76^{+0.017}_{-0.019}$ & $289^{+10.4}_{-10.5}$ & $0.828^{+0.0240}_{-0.0535}$ & 51111 \\ 
Kepler-1097 & $6.56^{+5.75}_{-6.52}$ & $-0.204^{+0.0716}_{-0.104}$ & $1.41^{+0.021}_{-0.018}$ & $59.1^{+10.7}_{-9.92}$ & $0.787^{+0.0331}_{-0.0417}$ & 51111 \\ 
Kepler-1143 & $2.56^{+8.18}_{-2.20}$ & $-0.177^{+0.0577}_{-0.0545}$ & $1.78^{+0.018}_{-0.019}$ & $0^{+0}_{-0}$ & $0.744^{+0.0130}_{-0.0304}$ & 52111 \\ 
Kepler-1318 & $8.45^{+3.16}_{-5.59}$ & $-0.162^{+0.0452}_{-0.0365}$ & $2.16^{+0.019}_{-0.02}$ & $175^{+12.7}_{-12.6}$ & $0.711^{+0.0115}_{-0.00909}$ & 52111 \\ 
Kepler-1362 & $1.4^{+2.61}_{-1.22}$ & $0.108^{+0.081}_{-0.0699}$ & $1.36^{+0.022}_{-0.022}$ & $128^{+15.2}_{-15.3}$ & $0.79^{+0.0185}_{-0.0165}$ & 31112 \\ 
Kepler-150 & $0.420^{+2.39}_{-0.350}$ & $0.0900^{+0.0385}_{-0.0419}$ & $1.11^{+0.017}_{-0.018}$ & $100^{+7.77}_{-7.75}$ & $0.981^{+0.0183}_{-0.0264}$ & 31111 \\  
Kepler-1512 & $5.38^{+7.90}_{-2.78}$ & $0.205^{+0.0362}_{-0.0263}$ & $5.38^{+0.066}_{-0.093}$ & $0^{+0}_{-0}$ & $0.747^{+0.0100}_{-0.00813}$ & 22211 \\
Kepler-1539 & $13.2^{+0.182}_{-0.386}$ & $0.374^{+0.0925}_{-0.122}$ & $0.621^{+0.018}_{-0.021}$ & $103^{+18.3}_{-18.1}$ & $1.26^{+0.137}_{-0.0272}$ & 42112 \\ 
Kepler-1544 & $1.44^{+1.12}_{-1.22}$ & $0.0252^{+0.0371}_{-0.0515}$ & $2.95^{+0.012}_{-0.011}$ & $114^{+6.19}_{-6.45}$ & $0.790^{+0.0107}_{-0.0109}$ & 31111 \\ 
Kepler-1545 & $11.99^{+1.08}_{-2.83}$ & $0.186^{+0.0516}_{-0.0495}$ & $1.37^{+0.020}_{-0.019}$ & $27.0^{+10.8}_{-11.6}$ & $0.890^{+0.0254}_{-0.0160}$ & 42111 \\ 
Kepler-1549 & $2.08^{+8.13}_{-2.01}$ & $0.239^{+0.0986}_{-0.0960}$ & $1.25^{+0.019}_{-0.022}$ & $58.9^{+16.6}_{-18.6}$ & $0.929^{+0.0205}_{-0.0398}$ & 52111 \\ 
Kepler-1554 & $8.48^{+3.72}_{-7.42}$ & $-0.431^{+0.0439}_{-0.0458}$ & $1.1^{+0.024}_{-0.025}$ & $234^{+26.8}_{-26.2}$ & $0.786^{+0.0232}_{-0.0256}$ & 51111 \\ 
Kepler-1593 & $0.488^{+1.22}_{-0.32}$ & $0.0524^{+0.0426}_{-0.0449}$ & $1.52^{+0.022}_{-0.023}$ & $264^{+14.7}_{-13.9}$ & $0.761^{+0.0109}_{-0.0109}$ & 31111 \\ 
Kepler-1600 & $7.6^{+4.23}_{-6.75}$ & $0.104^{+0.0471}_{-0.0462}$ & $1.01^{+0.025}_{-0.025}$ & $212^{+28.6}_{-30}$ & $0.879^{+0.0232}_{-0.0197}$ & 51111 \\ 
Kepler-1606 & $1.22^{+6.50}_{-1.16}$ & $-0.100^{+0.0707}_{-0.0894}$ & $1.19^{+0.019}_{-0.0118}$ & $97.2^{+11.5}_{-10.9}$ & $0.900^{+0.0191}_{-0.0352}$ & 31111 \\ 
Kepler-1634 & $1.07^{+5.23}_{-0.972}$ & $0.0493^{+0.0652}_{-0.0799}$ & $1.64^{+0.016}_{-0.017}$ & $150^{+8.03}_{-8.05}$ & $0.923^{+0.0173}_{-0.0310}$ & 31112 \\ 
Kepler-1635 & $11.8^{+1.21}_{-2.72}$ & $0.0694^{+0.0403}_{-0.0396}$ & $0.93^{+0.019}_{-0.021}$ & $54.3^{+12.6}_{-12}$ & $0.937^{+0.0061}_{-0.0475}$ & 41112 \\ 
Kepler-1636 & $9.34^{+0.912}_{-0.961}$ & $0.184^{+0.0489}_{-0.0484}$ & $0.524^{+0.023}_{-0.024}$ & $187^{+8.53}_{-8.46}$ & $1.06^{+0.016}_{-0.0209}$ & 11111 \\ 
Kepler-1690 & $2.21^{+6.36}_{-2.09}$ & $-0.101^{+0.0766}_{-0.0814}$ & $1.43^{+0.045}_{-0.049}$ & $134^{+7.58}_{-7.19}$ & $0.909^{+0.0234}_{-0.0274}$ & 31111 \\ 
Kepler-1701 & $1.20^{+4.91}_{-1.06}$ & $-0.0058^{+0.0516}_{-0.0624}$ & $1.73^{+0.016}_{-0.016}$ & $54.3^{+13.9}_{-14.0}$ & $0.870^{+0.0138}_{-0.0202}$ & 31111 \\ 
Kepler-1704 & $7.2^{+0.434}_{-0.482}$ & $0.196^{+0.0619}_{-0.0602}$ & $1.19^{+0.009}_{-0.009}$ & $68.3^{+27.4}_{-27.8}$ & $1.24^{+0.0238}_{-0.0141}$ & 11112 \\ 
Kepler-1708 & $12.4^{+0.805}_{-1.59}$ & $0.0622^{+0.0786}_{-0.0615}$ & $0.594^{+0.027}_{-0.027}$ & $213^{+24.4}_{-25.4}$ & $1.07^{+0.029}_{-0.0231}$ & 41111 \\ 
Kepler-174 & $12.3^{+0.745}_{-1.07}$ & $-0.340^{+0.0223}_{-0.0224}$ & $2.63^{+0.014}_{-0.013}$ & $6.16^{+4.01}_{-3.56}$ & $0.718^{+0.00529}_{-0.00234}$ & 42112 \\ 
Kepler-1840 & $12.31^{+0.899}_{-2.00}$ & $0.0150^{+0.0344}_{-0.0437}$ & $1.96^{+0.010}_{-0.02}$ & $152^{+9.58}_{-10.1}$ & $0.772^{+0.0105}_{-0.00853}$ & 42111 \\ 
Kepler-1868 & $0.894^{+2.02}_{-0.725}$ & $-0.0295^{0.0670}_{-0.0834}$ & $3.66^{+0.01}_{-0.01}$ & $119^{+7.93}_{-7.96}$ & $0.748^{+0.0104}_{-0.0153}$ & 31111 \\ 
Kepler-1981 & $4.96^{+1.89}_{-2.09}$ & $0.102^{+0.135}_{-0.135}$ & $1.53^{+0.011}_{-0.012}$ & $127^{+5.23}_{-5.25}$ & $1.10^{+0.0524}_{-0.0316}$ & 11112 \\ 
Kepler-22 & $0.498^{+1.87}_{-0.406}$ & $-0.196^{+0.0375}_{-0.0423}$ & $5.06^{+0.011}_{-0.011}$ & $69.8^{+18.8}_{-18}$ & $0.936^{+0.0113}_{-0.0289}$ & 31111 \\ 
Kepler-315 & $13.1^{+0.242}_{-0.525}$ & $-0.139^{+0.0352}_{-0.0365}$ & $0.925^{+0.02}_{-0.02}$ & $217^{+19.6}_{-18}$ & $0.93^{+0.00985}_{-0.0201}$ & 42221 \\
Kepler-419 & $2.54^{+0.107}_{-0.105}$ & $0.173^{+0.0502}_{-0.0462}$ & $0.994^{+0.012}_{-0.012}$ & $46.3^{+6.95}_{-7.84}$ & $1.40^{+0.0166}_{-0.0154}$ & 11111 \\ 
Kepler-424 & $11.7^{+1.06}_{-1.19}$ & $0.402^{+0.0453}_{-0.0416}$ & $1.43^{+0.015}_{-0.015}$ & $45.2^{+10.4}_{-10.1}$ & $1.02^{+0.00402}_{-0.00684}$ & 41111 \\ 
Kepler-442 & $6.93^{+4.15}_{-4.99}$ & $-0.387^{+0.0342}_{-0.0283}$ & $2.73^{+0.016}_{-0.015}$ & $128^{+9.57}_{-9.83}$ & $0.651^{+0.00851}_{-0.00608}$ & 51111 \\ 
Kepler-443 & $1.81^{+4.38}_{-1.59}$ & $0.0500^{+0.0415}_{-0.0433}$ & $1.25^{+0.023}_{-0.023}$ & $89.1^{+21.5}_{-21.5}$ & $0.789^{+0.0130}_{-0.0149}$ & 31111 \\ 
Kepler-452 & $5.12^{+1.99}_{-3.23}$ & $0.219^{+0.0915}_{-0.0880}$ & $1.81^{+0.01}_{-0.01}$ & $148^{+3.08}_{-3.06}$ & $1.06^{+0.0549}_{-0.0261}$ & 11111 \\ 
Kepler-51 & $0.154^{+0.466}_{-0.101}$ & $-0.00114^{+0.0314}_{-0.0343}$ & $1.24^{+0.015}_{-0.015}$ & $62.4^{+7.55}_{-7.73}$ & $0.975^{+0.0109}_{-0.0116}$ & 31111 \\ 
Kepler-553 & $10.1^{+2.30}_{-3.52}$ & $0.182^{+0.0436}_{-0.0395}$ & $1.35^{+0.019}_{-0.018}$ & $117^{+15.2}_{-15.7}$ & $0.924^{+0.0317}_{-0.0255}$ & 41111 \\ 
Kepler-56 & $6.72^{+1.27}_{-0.982}$ & $0.407^{+0.0462}_{-0.0473}$ & $1.08^{+0.012}_{-0.012}$ & $0^{+0}_{-0}$ & $1.26^{+0.0224}_{-0.0143}$ & 11112 \\ 
Kepler-62 & $11.7^{+1.23}_{-2.57}$ & $-0.278^{+0.0231}_{-0.0223}$ & $3.326^{+0.011}_{-0.011}$ & $0^{+0}_{-0}$ & $0.745^{+0.0178}_{-0.00721}$ & 42111 \\ 
Kepler-712 & $3.06^{+6.9}_{-2.85}$ & $-0.0973^{+0.0403}_{-0.0408}$ & $1.11^{+0.024}_{-0.026}$ & $145^{+13}_{-12.9}$ & $0.809^{+0.0158}_{-0.0197}$ & 51111 \\ 
Kepler-87 & $12.1^{+0.592}_{-0.559}$ & $-0.140^{+0.0576}_{-0.0501}$ & $0.781^{+0.015}_{-0.015}$ & $194^{+12.3}_{-13.3}$ & $1.46^{+0.00762}_{-0.180}$ & 11112 \\ 
Kepler-967 & $9.92^{+2.95}_{-8.43}$ & $-0.178^{+0.0678}_{-0.0578}$ & $2.1^{+0.01}_{-0.009}$ & $173^{+32.5}_{-32.9}$ & $0.801^{+0.0257}_{-0.0418}$ & 52112 \\ 
Kepler-97 & $3.55^{+2.61}_{-3.33}$ & $0.0312^{+0.0643}_{-0.0624}$ & $2.5^{+0.016}_{-0.016}$ & $288^{+21.9}_{-18.1}$ & $0.999^{+0.0410}_{-0.0257}$ & 31111 \\ 
KIC 10525077 & $8.07^{+2.68}_{-2.85}$ & $-0.0379^{+0.138}_{-0.137}$ & $0.67^{+0.022}_{-0.022}$ & $251^{+51.9}_{-50.7}$ & $1.1^{+0.0428}_{-0.0305}$ & 11111 \\ 
KIC 3526061 & $3.73^{+1.03}_{-0.547}$ & $0.0448^{+0.0822}_{-0.074}$ & $2.51^{+0.012}_{-0.012}$ & $170^{+1.68}_{-1.71}$ & $1.38^{+0.0903}_{-0.0735}$ & 11111 \\ 
PH2 & $1.44^{+2.24}_{-1.30}$ & $0.0257^{+0.0435}_{-0.0503}$ & $2.91^{+0.01}_{-0.01}$ & $3.63^{+2.94}_{-2.31}$ & $0.995^{+0.0263}_{-0.0315}$ & 31111 \\ 
TOI-1288 & $9.58^{+2.68}_{-8.47}$ & $0.124^{+0.0760}_{-0.0689}$ & $8.72^{+0.015}_{-0.013}$ & $52.3^{+11.9}_{-11.9}$ & $0.965^{+0.0496}_{-0.0264}$ & 41111 \\ 
TOI-1669 & $3.92^{+2.04}_{-2.77}$ & $0.266^{+0.0630}_{-0.0580}$ & $8.95^{+0.012}_{-0.012}$ & $54.7^{+18.4}_{-18.9}$ & $1.04^{+0.0406}_{-0.0291}$ & 11111 \\ 
TOI-1736 & $9.43^{+0.845}_{-0.818}$ & $0.123^{+0.058}_{-0.0569}$ & $11.3^{+0.014}_{-0.016}$ & $61.8^{+19.1}_{-18.5}$ & $1.13^{+0.00672}_{-0.00736}$ & 11111 \\ 
TOI-199 & $0.238^{+0.447}_{-0.174}$ & $0.177^{+0.0342}_{-0.0357}$ & $9.83^{+0.01}_{-0.011}$ & $44.6^{+5.22}_{-5.28}$ & $0.93^{+0.00757}_{-0.00989}$ & 31111 \\ 
TOI-2134 & $5.61^{+6.88}_{-3.07}$ & $0.201^{+0.0418}_{-0.0503}$ & $44.1^{+0.014}_{-0.014}$ & $0.148^{+0.127}_{-0.096}$ & $0.761^{+0.00446}_{-0.0125}$ & 21111 \\ 
TOI-2295 & $6.26^{+0.723}_{-1.03}$ & $0.346^{+0.0413}_{-0.0525}$ & $7.95^{+0.01}_{-0.01}$ & $10.1^{+5.74}_{-5.36}$ & $1.15^{+0.00928}_{-0.0138}$ & 11111 \\ 
TOI-4600 & $1.37^{+4.00}_{-1.24}$ & $0.157^{+0.0569}_{-0.0662}$ & $4.62^{+0.011}_{-0.012}$ & $23.6^{+8.17}_{-8.27}$ & $0.885^{+0.0168}_{-0.0241}$ & 31111 \\ 
TOI-712 & $1.96^{+3.7}_{-1.68}$ & $-0.14^{+0.0562}_{-0.0518}$ & $17^{+0.01}_{-0.009}$ & $37.1^{+29.6}_{-23.5}$ & $0.715^{+0.0106}_{-0.0101}$ & 31111 \\ 
WASP-41 & $1.31^{+1.63}_{-1.11}$ & $0.041^{+0.00969}_{-0.0102}$ & $6.12^{+0.02}_{-0.021}$ & $74.4^{+5.07}_{-5.11}$ & $0.952^{+0.0183}_{-0.018}$ & 31111 \\ 
WASP-47 & $7.73^{+1.02}_{-0.991}$ & $0.413^{+0.0485}_{-0.0493}$ & $3.7^{+0.021}_{-0.02}$ & $29.3^{+6.4}_{-6.5}$ & $1.04^{+0.00891}_{-0.00587}$ & 11111 \\ 

\end{longtable}

%% file: HZ_Residency.txt
\begin{ThreePartTable}
\begin{TableNotes}[flushleft]
\footnotesize
\item \textit{Note.} The 9 planets found to be just outside the HZ have
dashes under the HZ Membership column. The time in the HZ is the
continuous time either since the planet most recently entered the
HZ until the present time, or from the 30 Myr lower limit until the
present time. If the planet was outside the HZ at any time, the
time spent in the HZ refers to the time since its most recent entry.
\end{TableNotes}

\begin{longtable*}{lccccc}
\caption{Planetary HZ Residency} \label{tab:HZresidency} \\

\toprule
Planet & HZ Inner (AU) & HZ Outer (AU) & Orbital Distance (AU) & HZ Membership & Time in HZ (Gyr) \\
\midrule
\endfirsthead

\toprule
Planet & HZ Inner (AU) & HZ Outer (AU) & Orbital Distance (AU) & HZ Membership & Time in HZ (Gyr) \\
\midrule
\endhead

\midrule
\multicolumn{6}{r}{{Continued on next page}} \\
\endfoot

\bottomrule
\endlastfoot
 
BD+14 4559 b & 0.538 & 1.33 & 0.78 & Continuous & 11.6 \\ 
BD+55 362 b & 0.35 & 0.857 & 0.78 & Continuous & 9.50 \\ 
BD-08 2823 c & 0.296 & 0.734 & 0.68 & Continuous & 1.89 \\ 
HAT-P-13 c & 1.12 & 2.67 & 1.26 & Continuous & 4.40 \\ 
HD 100777 b & 0.775 & 1.85 & 1.03 & Continuous & 3.41 \\ 
HD 10180 g & 0.911 & 2.14 & 1.43 & Continuous & 5.25 \\ 
HD 103891 b & 1.83 & 4.28 & 3.27 & Cold Start & 1.61 \\ 
HD 103949 b & 0.305 & 0.756 & 0.439 & Continuous & 4.58 \\ 
HD 10442 b & 1.1 & 2.7 & 2.34 & Cold Start & 0.540 \\ 
HD 106270 b & 1.81 & 4.33 & 3.34 & Cold Start & 2.64 \\ 
HD 108874 b & 0.757 & 1.81 & 1.05 & Continuous & 7.90 \\ 
HD 109286 b & 0.796 & 1.89 & 1.26 & Continuous & 0.405 \\ 
HD 111998 b & 1.36 & 3.11 & 1.91 & Continuous & 1.44 \\ 
HD 114783 b & 0.502 & 1.22 & 1.2 & Cold Start & 2.01 \\ 
HD 11506 b & 1.09 & 2.56 & 2.43 & Cold Start & 0.765 \\ 
HD 124330 b & 0.502 & 1.18 & 0.86 & Continuous & 6.33 \\ 
HD 128311 b & 0.443 & 1.09 & 1.09 & Cold Start & 1.96 \\ 
HD 128356 b & 0.467 & 1.15 & 0.87 & Continuous & 1.77 \\ 
HD 13167 b & 1.74 & 4.12 & 4.1 & - & - \\ 
HD 132406 b & 1.01 & 2.39 & 1.98 & Cold Start & 2.20 \\ 
HD 134606 d & 0.815 & 1.94 & 1.94 & Cold Start & 1.06 \\ 
HD 13724 c & 0.468 & 1.11 & 0.534 & Dessicated & 0.176 \\ 
HD 137388 b & 0.52 & 1.26 & 0.89 & Continuous & 0.620 \\ 
HD 137496 c & 1.2 & 2.84 & 1.22 & - & - \\ 
HD 13908 c & 1.47 & 3.41 & 2.03 & Continuous & 3.83 \\ 
HD 141399 d & 0.953 & 2.27 & 2.09 & - & - \\ 
HD 141937 b & 0.809 & 1.9 & 1.5 & Continuous & 0.266 \\ 
HD 142415 b & 0.796 & 1.86 & 1.05 & Continuous & 0.247 \\ 
HD 145934 b & 2.92 & 7.23 & 4.6 & Cold Start & 0.421 \\ 
HD 153950 b & 1.1 & 2.57 & 1.28 & Continuous & 6.06 \\ 
HD 155193 b & 0.57 & 1.32 & 1.04 & Continuous & 3.32 \\ 
HD 155358 c & 0.47 & 1.1 & 1.02 & Cold Start & 6.17 \\ 
HD 156411 b & 1.73 & 4.07 & 1.88 & Continuous & 4.00 \\ 
HD 159868 b & 1.44 & 3.43 & 2.25 & Cold Start & 3.33 \\ 
HD 1605 c & 2.01 & 4.98 & 3.52 & Cold Start & 1.72 \\  
HD 16175 b & 1.33 & 3.12 & 2.15 & Continuous & 3.60 \\  
HD 163607 c & 1.23 & 2.94 & 2.39 & Cold Start & 0.661 \\ 
HD 165155 b & 0.637 & 1.53 & 1.13 & Continuous & 1.82 \\ 
HD 169830 c & 1.58 & 3.64 & 3.08 & Cold Start & 2.24 \\ 
HD 175167 b & 1.26 & 2.99 & 2.4 & Cold Start & 1.86 \\ 
HD 17674 b & 0.92 & 2.16 & 1.42 & Continuous & 8.32 \\ 
HD 18015 b & 2.22 & 5.3 & 3.87 & Cold Start & 1.38 \\ 
HD 181720 b & 1.05 & 2.48 & 1.85 & Cold Start & 3.81 \\ 
HD 183263 b & 0.994 & 2.33 & 1.49 & Continuous & 3.49 \\ 
HD 190647 b & 1.06 & 2.53 & 2.07 & Cold Start & 4.03 \\ 
HD 191939 g & 0.616 & 1.48 & 0.8 & Continuous & 12.6 \\ 
HD 2039 b & 0.939 & 2.2 & 2.2 & - & - \\ 
HD 208487 b & 0.495 & 1.16 & 0.51 & - & - \\ 
HD 20868 b & 0.424 & 1.05 & 0.95 & Continuous & 2.80 \\ 
HD 210277 b & 0.73 & 1.75 & 1.13 & Continuous & 8.23 \\ 
HD 216437 b & 1.12 & 2.62 & 2.5 & Cold Start & 0.001 \\ 
HD 216520 c & 0.458 & 1.12 & 0.528 & Desiccated & 1.03 \\ 
HD 218566 b & 0.465 & 1.15 & 0.69 & Continuous & 11.1 \\ 
HD 219415 b & 1.59 & 3.94 & 3.2 & Cold Start & 0.923 \\ 
HD 221287 b & 0.947 & 2.19 & 1.25 & Continuous & 0.446 \\ 
HD 221585 b & 1.23 & 2.92 & 2.31 & Cold Start & 3.35 \\ 
HD 224538 b & 1.27 & 2.97 & 2.44 & Cold Start & 2.63 \\ 
HD 23079 b & 0.852 & 2 & 1.5 & Continuous & 3.50 \\ 
HD 23127 b & 1.14 & 2.69 & 2.37 & Cold Start & 1.34 \\ 
HD 24040 c & 1.04 & 2.46 & 1.3 & Continuous & 7.09 \\ 
HD 27969 b & 1.01 & 2.36 & 1.55 & Continuous & 5.67 \\ 
HD 28185 b & 0.77 & 1.83 & 1.04 & Continuous & 2.29 \\ 
HD 33564 b & 0.576 & 1.33 & 1.1 & Continuous & 1.77 \\ 
HD 34445 b & 1.06 & 2.5 & 2.07 & Cold Start & 2.26 \\ 
HD 34445 f & 1.06 & 2.5 & 1.54 & Continuous & 7.76 \\ 
HD 38801 b & 1.49 & 3.59 & 1.62 & Continuous & 4.58 \\ 
HD 40307 g & 0.396 & 0.978 & 0.607 & Continuous & 2.10  \\ 
HD 4203 b & 1.06 & 2.52 & 1.17 & Continuous & 6.93 \\ 
HD 43197 b & 0.648 & 1.55 & 0.882 & Continuous & 4.18 \\ 
HD 44219 b & 1.01 & 2.4 & 1.26 & Continuous & 9.62 \\ 
HD 45350 b & 0.89 & 2.12 & 1.92 & Cold Start & 0.172 \\ 
HD 45364 b & 0.607 & 1.45 & 0.679 & Continuous & 7.54 \\ 
HD 45364 c & 0.607 & 1.45 & 0.902 & Continuous & 7.54 \\ 
HD 48265 b & 1.48 & 3.51 & 1.81 & Continuous & 4.66 \\ 
HD 48948 d & 0.337 & 0.843 & 0.489 & Continuous & 1.13 \\ 
HD 564 b & 0.787 & 1.85 & 1.2 & Continuous & 0.573 \\ 
HD 63765 b & 0.558 & 1.34 & 0.94 & Continuous & 1.24 \\ 
HD 7199 b & 0.638 & 1.54 & 1.36 & Continuous & 3.35 \\ 
HD 73526 c & 0.474 & 1.12 & 1.03 & Continuous & 0.019 \\ 
HD 73534 b & 1.45 & 3.56 & 2.95 & Cold Start & 0.820 \\ 
HD 82943 b & 0.933 & 2.19 & 1.19 & Continuous & 0.246 \\ 
HD 8326 b & 0.323 & 0.797 & 0.533 & Continuous & 3.96 \\ 
HD 86264 b & 1.58 & 3.66 & 2.76 & Continuous & 1.65 \\ 
HD 9174 b & 1.17 & 2.8 & 2.2 & Cold Start & 3.31 \\ 
HD 92788 b & 0.767 & 1.81 & 0.96 & Continuous & 4.60 \\ 
HD 94834 b & 2.26 & 5.59 & 2.74 & Continuous & 3.55 \\ 
HD 95089 c & 2.86 & 7.04 & 3.33 & Continuous & 2.53 \\ 
HD 99109 b & 0.64 & 1.55 & 1.12 & Continuous & 1.09 \\ 
HIP 5158 b & 0.409 & 1.01 & 0.89 & Cold Start & 6.51 \\ 
HIP 56640 b & 2.62 & 6.48 & 3.77 & Cold Start & 1.75 \\ 
HIP 57274 d & 0.976 & 2.44 & 1.01 & Continuous & 11.9 \\ 
K2-312 c & 0.967 & 2.27 & 1.96 & Cold Start & 2.45 \\ 
kap CrB b & 2.8 & 6.9 & 2.8 & Cold Start & 2.56 \\ 
KELT-6 c & 1.33 & 3.07 & 2.39 & Cold Start & 2.87 \\ 
Kepler-1038 b & 0.506 & 1.23 & 0.515 & - & - \\ 
Kepler-1097 b & 0.48 & 1.16 & 0.579 & Continuous & 6.50 \\ 
Kepler-1143 c & 0.38 & 0.94 & 0.648 & Continuous & 2.50 \\ 
Kepler-1318 b & 0.34 & 0.856 & 0.609 & Continuous & 8.39 \\ 
Kepler-1362 b & 0.39 & 0.967 & 0.457 & Continuous & 1.39 \\ 
Kepler-150 f & 0.63 & 1.51 & 1.24 & Continuous & 0.377 \\ 
Kepler-1512 b & 0.111 & 0.278 & 0.129 & Desiccated & 5.07 \\
Kepler-1539 b & 0.473 & 1.16 & 0.478 & Continuous & 13.1 \\ 
Kepler-1544 b & 0.401 & 0.989 & 0.54 & Continuous & 1.39 \\ 
Kepler-1545 b & 0.56 & 1.37 & 0.563 & - & - \\ 
Kepler-1549 b & 0.558 & 1.35 & 0.669 & Continuous & 2.03 \\ 
Kepler-1554 b & 0.483 & 1.16 & 0.626 & Continuous & 8.44 \\ 
Kepler-1593 b & 0.353 & 0.883 & 0.555 & Continuous & 0.465 \\ 
Kepler-1600 b & 0.519 & 1.26 & 1.01 & Continuous & 7.52 \\ 
Kepler-1606 b & 0.624 & 1.5 & 0.642 & Desiccated & 1.17 \\ 
Kepler-1634 b & 0.556 & 1.32 & 1.05 & Continuous & 1.03 \\ 
Kepler-1635 b & 0.55 & 1.32 & 1.11 & Desiccated & 10.7 \\ 
Kepler-1636 b & 0.978 & 2.29 & 1.15 & Continuous & 9.28 \\ 
Kepler-1690 b & 0.576 & 1.38 & 0.71 & Continuous & 2.17 \\ 
Kepler-1701 b & 0.509 & 1.23 & 0.569 & Desiccated & 1.14 \\ 
Kepler-1704 b & 1.26 & 2.99 & 2.01 & Cold Start & 4.76 \\ 
Kepler-1708 b & 0.912 & 2.12 & 1.64 & Cold Start & 6.30 \\ 
Kepler-174 d & 0.403 & 0.995 & 0.677 & Continuous & 12.2 \\ 
Kepler-1840 b & 0.398 & 0.992 & 0.447 & Continuous & 12.1 \\ 
Kepler-1868 b & 0.366 & 0.914 & 0.611 & Continuous & 0.825 \\ 
Kepler-1981 b & 0.99 & 2.33 & 1.17 & Continuous & 4.93 \\ 
Kepler-22 b & 0.606 & 1.44 & 0.849 & Continuous & 0.460 \\ 
Kepler-315 c & 0.757 & 1.8 & 0.761 & - & - \\ 
Kepler-419 c & 1.58 & 3.63 & 1.68 & Continuous & 2.50 \\ 
Kepler-424 c & 0.64 & 1.53 & 0.73 & Continuous & 11.7 \\ 
Kepler-442 b & 0.304 & 0.761 & 0.393 & Continuous & 6.86 \\ 
Kepler-443 b & 0.407 & 1.01 & 0.572 & Continuous & 1.74 \\ 
Kepler-452 b & 0.829 & 1.96 & 0.994 & Continuous & 5.06 \\ 
Kepler-51 e & 0.632 & 1.5 & 0.791$*$ & Continuous & 0.114 \\ 
Kepler-553 c & 0.561 & 1.34 & 0.917 & Continuous & 10.1 \\ 
Kepler-56 d & 2.29 & 5.66 & 2.31$*$ & Cold Start & 4.12 \\ 
Kepler-62 e & 0.400 & 0.988 & 0.428 & Continuous & 11.6 \\
Kepler-62 f & 0.400 & 0.988 & 0.720 & Continuous & 11.6 \\
Kepler-712 c & 0.443 & 1.09 & 0.682 & Continuous & 3.14 \\ 
Kepler-87 c & 1.08 & 2.54 & 1.34 & Desiccated & 11.7 \\ 
Kepler-967 c & 0.375 & 0.915 & 0.568 & Continuous & 9.87 \\ 
Kepler-97 c & 0.736 & 1.74 & 1.67 & Desiccated & 1.59 \\
KIC 10525077 b & 0.832 & 1.94 & 1.76$*$ & Cold Start & 1.46 \\ 
KIC 3526061 b & 3.15 & 7.8 & 5.14 & - & - \\ 
PH2 b & 0.71 & 1.69 & 0.826 & Continuous & 1.40 \\ 
TOI-1288 c & 0.633 & 1.53 & 1.07 & Continuous & 9.53 \\ 
TOI-1669 b & 0.728 & 1.74 & 1.27 & Continuous & 3.87 \\  
TOI-1736 c & 1.03 & 2.45 & 1.37 & Continuous & 9.37 \\ 
TOI-199 c & 0.519 & 1.26 & 0.807 & Continuous & 0.189 \\ 
TOI-2134 c & 0.344 & 0.862 & 0.371 & Continuous & 5.55 \\ 
TOI-2295 c & 1.08 & 2.56 & 2.02 & Cold Start & 4.66 \\ 
TOI-4600 c & 0.5 & 1.22 & 1.15 & Continuous & 1.32 \\ 
TOI-712 d & 0.339 & 0.848 & 0.341 & Desiccated & 1.87 \\ 
WASP-41 c & 0.607 & 1.45 & 1.07 & Continuous & 1.29 \\ 
WASP-47 c & 0.81 & 1.93 & 1.36 & Continuous & 7.70 

\\
\bottomrule
\insertTableNotes

\end{longtable*}
\end{ThreePartTable}

%% file: PriorTable.txt
\begin{ThreePartTable}

\begin{TableNotes}[flushleft]
\footnotesize
\item \textit{Note.} Dust extinction values are constrained to be non-negative. For cases where the adopted mean extinction is zero, uncertainties are therefore reported as upper uncertainties only (i.e., $A_G = 0^{+\,\sigma}$) rather than as symmetric $\pm$ values.
a: \cite{Rosenthal2021}; b: \cite{Sousa2021}; c: \cite{Dalal2021}; d: \cite{Sousa2024}; e: \cite{Demangeon2021}; f: \cite{Borgniet2017}; g: \cite{Silva2022}; j: \cite{Dalal2024}; k: \cite{Teng2023}; l: \cite{Morton2016}; m: \cite{Petigura2018}; n: \cite{Torres2017}; o: \cite{Latham2005}; p: \cite{Armstrong2021}; q: \cite{Dalba2021}; r: \cite{Kipping2022}; s: \cite{Valizadegan2022}; t: \cite{Rowe2014}; u: \cite{Torres2015}; v: \cite{Jenkins2015}; w: \cite{Ofir2014}; x: \cite{Wang2015}; y: \cite{Karjalainen2022}; z: \cite{Dalba2024}; aa: \cite{Knudstrup2023}; ab: \cite{VanZandt2023}; ac: \cite{Polanski2024}; ad: \cite{Hobson2023}; ae: \cite{Rescigno2024}; af: \cite{Heidari2025}; ag: \cite{Mireles2023}; ah: \cite{Thompson2018}
\end{TableNotes}

\begin{longtable*}{lcccc}
\caption{Priors Used in This Study} \label{tab:priors} \\

\toprule
Host Star & [Fe/H] & Reference & Parallax (mas) & $A_{g}$ (mmag) \\
\midrule
\endfirsthead

\toprule
Host Star & [Fe/H] & Reference & Parallax (mas) & $A_{g}$ (mmag) \\
\midrule
\endhead

\midrule
\multicolumn{5}{r}{{Continued on next page}} \\
\endfoot

\bottomrule
\endlastfoot

BD-08 2823 & $0.01\pm0.11$ & \tnote{b} & $24.18\pm0.014$ & $0.2\pm0.35$ \\ 
BD+14 4559 & $0.21\pm0.05$ & \tnote{b} & $20.28\pm0.015$ & $0.2\pm1.05$ \\ 
BD+55 362 & $0.18\pm0.05$ & \tnote{c} & $19.03\pm0.014$ & $1.6\pm8.65$ \\ 
HAT-P-13 & $0.48\pm0.03$ & \tnote{b} & $4.075\pm0.019$ & $35.1\pm4.25$ \\ 
HD 100777 & $0.28\pm0.02$ & \tnote{b} & $20.16\pm0.029$ & $0.1\pm0.3$ \\ 
HD 10180 & $0.08\pm0.01$ & \tnote{b} & $25.66\pm0.015$ & $0\pm0.05$ \\ 
HD 103891 & $-0.166\pm0.014$ & \tnote{d} & $18.22\pm0.039$ & $0\pm0$ \\ 
HD 103949 & $-0.05\pm0.04$ & \tnote{b} & $37.74\pm0.029$ & $0\pm0.05$ \\ 
HD 10442 & $0\pm0.03$ & \tnote{b} & $7.508\pm0.028$ & $134.1\pm1.25$ \\ 
HD 106270 & $0.11\pm0.02$ & \tnote{b} & $10.53\pm0.031$ & $316.2\pm7$ \\ 
HD 108874 & $0.21\pm0.02$ & \tnote{b} & $16.73\pm0.24$ & $0.1\pm0.2$ \\ 
HD 109286 & $0.05\pm0.02$ & \tnote{e} & $18.1\pm0.025$ & $0.4\pm3.45$ \\ 
HD 111998 & $0.07\pm0.1$ & \tnote{f} & $29.91\pm0.038$ & $0\pm0.05$ \\ 
HD 114783 & $0.04\pm0.04$ & \tnote{b} & $47.55\pm0.029$ & $0\pm0.1$ \\ 
HD 11506 & $0.34\pm0.02$ & \tnote{b} & $19.53\pm0.022$ & $0\pm0.05$ \\ 
HD 124330 & $0.22\pm0.01$ & \tnote{c} & $16.63\pm0.015$ & $0.1\pm0.1$ \\ 
HD 128311 & $-0.03\pm0.03$ & \tnote{b} & $61.28\pm0.043$ & $0\pm0.05$ \\ 
HD 128356 & $0.26\pm0.06$ & \tnote{b} & $38.38\pm0.028$ & $0.1\pm0.25$ \\ 
HD 13167 & $0.245\pm0.016$ & \tnote{d} & $6.746\pm0.022$ & $210.3\pm3.85$ \\ 
HD 132406 & $0.12\pm0.01$ & \tnote{b} & $14.18\pm0.019$ & $0.2\pm0.35$ \\ 
HD 134606 & $0.31\pm0.03$ & \tnote{b} & $37.3\pm0.018$ & $0.5\pm0.5$ \\ 
HD 13724 & $0.22\pm0.02$ & \tnote{b} & $23.02\pm0.01$8 & $0\pm0.05$ \\ 
HD 137388 & $0.2\pm0.04$ & \tnote{b} & $24.62\pm0.013$ & $0\pm0.05$ \\ 
HD 137496 & $-0.03\pm0.04$ & \tnote{g} & $6.422\pm0.019$ & $200.2\pm26.6$ \\ 
HD 13908 & $0\pm0.02$ & \tnote{b} & $12.52\pm0.02$ & $0.2\pm0.3$ \\ 
HD 141399 & $0.3\pm0.04$ & \tnote{b} & $26.99\pm0.015$ & $0.2\pm0.6$ \\ 
HD 141937 & $0.14\pm0.01$ & \tnote{b} & $30.56\pm0.09$ & $0\pm0.1$ \\ 
HD 142415 & $0.14\pm0.02$ & \tnote{b} & $28.21\pm0.011$ & $0\pm0.1$ \\   
HD 145934 & $0.095\pm0.06$ & \tnote{a} & $4.372\pm0.022$ & $20.8\pm2.25$ \\ 
HD 153950 & $-0.01\pm0.01$ & \tnote{b} & $20.66\pm0.025$ & $0\pm0.1$ \\ 
HD 155193 & $0.05\pm0.01$ & \tnote{c} & $17.33\pm0.02$ & $4.4\pm5.95$ \\ 
HD 155358 & $-0.64\pm0.02$ & \tnote{b} & $22.92\pm0.013$ & $0.2\pm0.35$ \\ 
HD 156411 & $-0.1\pm0.01$ & \tnote{b} & $17.97\pm0.021$ & $36.6\pm2.95$ \\ 
HD 159868 & $-0.04\pm0.01$ & \tnote{b} & $17.88\pm0.021$ & $94.8\pm13.7$ \\ 
HD 1605 & $0.13\pm0.03$ & \tnote{b} & $11.36\pm0.021$ & $75.5\pm3.35$ \\  
HD 16175 & $0.29\pm0.02$ & \tnote{b} & $16.67\pm0.028$ & $0.1\pm0.15$ \\ 
HD 163607 & $0.22\pm0.02$ & \tnote{b} & $14.76\pm0.018$ & $4.5\pm13.8$ \\ 
HD 165155 & $0.12\pm0.02$ & \tnote{b} & $15.78\pm0.022$ & $33.1\pm3.7$ \\ 
HD 169830 & $0.18\pm0.02$ & \tnote{b} & $27.15\pm0.15$ & $0\pm0$ \\ 
HD 175167 & $0.35\pm0.07$ & \tnote{b} & $14.04\pm0.023$ & $80.5\pm15.7$ \\ 
HD 17674 & $-0.17\pm0.01$ & \tnote{b} & $22.52\pm0.025$ & $0\pm0.05$ \\ 
HD 18015 & $-0.08\pm0.013$ & \tnote{d} & $8.013\pm0.022$ & $0\pm0$ \\ 
HD 181720 & $-0.51\pm0.02$ & \tnote{b} & $16.68\pm0.027$ & $39.8\pm5.6$ \\ 
HD 183263 & $0.31\pm0.02$ & \tnote{b} & $18.34\pm0.021$ & $0.1\pm0.15$ \\ 
HD 190647 & $0.24\pm0.02$ & \tnote{b} & $18.39\pm0.025$ & $0.6\pm2.6$ \\ 
HD 191939 & $-0.22\pm0.02$ & \tnote{b} & $18.7\pm0.013$ & $0.1\pm0.15$ \\ 
HD 2039 & $0.33\pm0.02$ & \tnote{b} & $11.77\pm0.014$ & $19\pm4.85$ \\ 
HD 208487 & $0.09\pm0.01$ & \tnote{b} & $22.27\pm0.03$ & $0\pm0.1$ \\ 
HD 20868 & $0.07\pm0.04$ & \tnote{b} & $20.96\pm0.013$ & $0.8\pm1.4$ \\ 
HD 210277 & $0.19\pm0.02$ & \tnote{b} & $46.85\pm0.028$ & $0\pm0.05$ \\ 
HD 216437 & $0.24\pm0.02$ & \tnote{b} & $37.46\pm0.025$ & $0\pm0.05$ \\ 
HD 216520 & $-0.34\pm0.02$ & \tnote{d} & $51.16\pm0.018$ & $0\pm0$ \\ 
HD 218566 & $0.21\pm0.06$ & \tnote{b} & $34.7\pm0.029$ & $0\pm0.05$ \\ 
HD 219415 & $0.01\pm0.03$ & \tnote{b} & $6.044\pm0.013$ & $165.5\pm1$ \\ 
HD 221287 & $0.07\pm0.02$ & \tnote{b} & $17.87\pm0.021$ & $0\pm0.1$ \\ 
HD 221585 & $0.29\pm0.02$ & \tnote{b} & $17.92\pm0.016$ & $0\pm0.1$ \\ 
HD 224538 & $0.37\pm0.02$ & \tnote{b} & $12.64\pm0.02$ & $0.4\pm0.45$ \\ 
HD 23079 & $-0.12\pm0.01$ & \tnote{b} & $29.86\pm0.019$ & $0\pm0.05$ \\ 
HD 23127 & $0.38\pm0.02$ & \tnote{b} & $10.69\pm0.014$ & $24.3\pm1.65$ \\ 
HD 24040 & $0.23\pm0.01$ & \tnote{b} & $21.47\pm0.024$ & $0.2\pm0.5$ \\ 
HD 27969 & $0.18\pm0.02$ & \tnote{e} & $14.78\pm0.028$ & $0.6\pm1.1$ \\ 
HD 28185 & $0.21\pm0.02$ & \tnote{b} & $25.49\pm0.021$ & $0.3\pm0.65$ \\ 
HD 33564 & $0.253\pm0.048$ & \tnote{d} & $48.11\pm0.073$ & $0\pm0.05$ \\ 
HD 34445 & $0.16\pm0.02$ & \tnote{b} & $21.84\pm0.021$ & $0.2\pm0.3$ \\ 
HD 38801 & $0.26\pm0.03$ & \tnote{b} & $11.03\pm0.02$2 & $0\pm0$ \\ 
HD 40307 & $-0.34\pm0.04$ & \tnote{b} & $77.33\pm0.017$ & $0\pm0$ \\ 
HD 4203 & $0.38\pm0.03$ & \tnote{b} & $12.3\pm0.02$ & $69.3\pm2.65$ \\ 
HD 43197 & $0.4\pm0.03$ & \tnote{b} & $16.02\pm0.011$ & $0.5\pm0.95$ \\ 
HD 44219 & $0.06\pm0.01$ & \tnote{b} & $18.9\pm0.019$ & $0.5\pm0.5$ \\ 
HD 45350 & $0.28\pm0.02$ & \tnote{b} & $21.29\pm0.029$ & $0.9\pm1.6$ \\ 
HD 45364 & $-0.14\pm0.02$ & \tnote{b} & $29.13\pm0.017$ & $5.2\pm4.1$ \\ 
HD 48265 & $0.39\pm0.03$ & \tnote{b} & $11.01\pm0.016$ & $0\pm0$ \\ 
HD 48948 & $-0.21\pm0.03$ & \tnote{j} & $59.39\pm0.025$ & $0\pm0$ \\ 
HD 564 & $-0.19\pm0.01$ & \tnote{b} & $19.54\pm0.018$ & $0.1\pm0.35$ \\ 
HD 63765 & $-0.14\pm0.03$ & \tnote{b} & $30.7\pm0.018$ & $0.1\pm0.2$ \\  
HD 7199 & $0.33\pm0.03$ & \tnote{b} & $27.69\pm0.016$ & $0\pm0.25$ \\ 
HD 73526 & $0.26\pm0.01$ & \tnote{b} & $10.33\pm0.022$ & $72\pm17.3$ \\ 
HD 73534 & $0.19\pm0.04$ & \tnote{b} & $11.96\pm0.025$ & $78\pm20$ \\ 
HD 80653 & $0.333\pm0.018$ & \tnote{d} & $9.26\pm0.022$ & $4.5\pm4.45$ \\ 
HD 82943 & $0.27\pm0.02$ & \tnote{b} & $36.12\pm0.026$ & $0\pm0.05$ \\ 
HD 8326 & $0.04\pm0.03$ & \tnote{b} & $32.56\pm0.022$ & $0\pm0.05$ \\ 
HD 86264 & $0.41\pm0.05$ & \tnote{b} & $14.86\pm0.021$ & $0.3\pm0.4$ \\ 
HD 9174 & $0.39\pm0.03$ & \tnote{b} & $12.3\pm0.022$ & $110.3\pm1.85$ \\ 
HD 92788 & $0.29\pm0.02$ & \tnote{b} & $28.95\pm0.026$ & $0\pm0.05$ \\ 
HD 94834 & $0.048\pm0.028$ & \tnote{d} & $10.29\pm0.027$ & $141.2\pm2.95$ \\ 
HD 95089 & $0.03\pm0.04$ & \tnote{b} & $7.378\pm0.025$ & $97.3\pm4.8$ \\ 
HD 99109 & $0.34\pm0.04$ & \tnote{b} & $18.18\pm0.017$ & $0.2\pm0.8$ \\ 
HIP 5158 & $0.22\pm0.07$ & \tnote{b} & $19.32\pm0.02$ & $0\pm0.05$ \\ 
HIP 56640 & $0.128\pm0.037$ & \tnote{d} & $8.223\pm0.02$ & $190.6\pm2.3$ \\ 
HIP 57274 & $-0.07\pm0.05$ & \tnote{b} & $38.68\pm0.024$ & $0.1\pm0.2$ \\ 
kap CrB & $0.09\pm0.085$ & \tnote{k} & $33.34\pm0.08$ & $0.1\pm0.15$ \\ 
KELT-6 & $-0.23\pm0.02$ & \tnote{b} & $4.106\pm0.018$ & $5.6\pm11.9$ \\ 
Kepler-1038 & $-0.04\pm0.17$ & \tnote{l} & $1.749\pm0.018$ & $285.9\pm10.5$ \\ 
Kepler-1058 & $-0.06\pm0.17$ & \tnote{l} & $1.401\pm0.076$ & $796\pm8.65$ \\ 
Kepler-1097 & $-0.11\pm0.17$ & \tnote{l} & $1.404\pm0.02$ & $54.2\pm11$ \\ 
Kepler-1143 & $-0.06\pm0.16$ & \tnote{l} & $1.77\pm0.019$ & $0\pm0$ \\  
Kepler-1318 & $-0.01\pm0.14$ & \tnote{l} & $2.15\pm0.021$ & $171.6\pm12.4$ \\ 
Kepler-1362 & $0.06\pm0.15$ & \tnote{l} & $1.362\pm0.025$ & $125.3\pm15.6$ \\ 
Kepler-150 & $0.12\pm0.04$ & \tnote{m} & $1.103\pm0.019$ & $97.8\pm7.95$ \\ 
Kepler-1512 & $0.1\pm0.04$ & \tnote{m} & $3.567\pm0.14$ & $0\pm0$ \\ 
Kepler-1539 & $-0.01\pm0.16$ & \tnote{l} & $1.314\pm0.018$ & $84\pm16.6$ \\ 
Kepler-1544 & $0.01\pm0.04$ & \tnote{m} & $2.954\pm0.011$ & $112.8\pm6.7$ \\ 
Kepler-1545 & $-0.06\pm0.15$ & \tnote{l} & $1.347\pm0.02$ & $18.2\pm11.4$ \\ 
Kepler-1549 & $0.02\pm0.15$ & \tnote{l} & $1.239\pm0.02$ & $50.8\pm17.6$ \\ 
Kepler-1554 & $-0.46\pm0.04$ & \tnote{m} & $1.074\pm0.026$ & $222.4\pm28.7$ \\ 
Kepler-1593 & $0.07\pm0.04$ & \tnote{m} & $1.545\pm0.027$ & $265.6\pm14.7$ \\ 
Kepler-1600 & $0.11\pm0.04$ & \tnote{m} & $0.9813\pm0.028$ & $175.8\pm29.6$ \\ 
Kepler-1606 & $0.01\pm0.15$ & \tnote{l} & $1.186\pm0.02$ & $92.8\pm11.8$ \\ 
Kepler-1634 & $0.02\pm0.15$ & \tnote{l} & $1.634\pm0.017$ & $147\pm8.05$ \\ 
Kepler-1635 & $0.06\pm0.04$ & \tnote{m} & $0.888\pm0.026$ & $38\pm13.3$ \\ 
Kepler-1636 & $0.19\pm0.04$ & \tnote{m} & $0.5177\pm0.024$ & $184.7\pm8.65$ \\ 
Kepler-1653 & $-0.18\pm0.1$ & \tnote{n} & $1.109\pm0.049$ & $0\pm0$ \\ 
Kepler-1690 & $-0.062\pm0.1$ & \tnote{o} & $1.404\pm0.05$ & $133.3\pm7.5$ \\ 
Kepler-1701 & $0.03\pm0.1$ & \tnote{p} & $1.728\pm0.016$ & $49.5\pm14.3$ \\ 
Kepler-1704 & $0.2\pm0.06$ & \tnote{q} & $1.187\pm0.0094$ & $61.4\pm26.9$ \\ 
Kepler-1708 & $0\pm0.2$ & \tnote{r} & $0.5893\pm0.026$ & $163.7\pm27.8$ \\ 
Kepler-174 & $-0.56\pm0.04$ & \tnote{m} & $2.601\pm0.014$ & $3.9\pm4.2$ \\ 
Kepler-1840 & $0.14\pm0.15$ & \tnote{s} & $1.939\pm0.02$ & $148.2\pm9.75$ \\ 
Kepler-1868 & $0\pm0.15$ & \tnote{s} & $3.662\pm0.011$ & $118.7\pm8$ \\ 
Kepler-1981 & $0.12\pm0.15$ & \tnote{ah} & $1.53\pm0.012$ & $126.7\pm5.35$ \\ 
Kepler-22 & $-0.2\pm0.04$ & \tnote{m} & $5.063\pm0.011$ & $67.3\pm19.2$ \\ 
Kepler-315 & $-0.27\pm0.04$ & \tnote{m} & $0.8742\pm0.021$ & $149.4\pm20.5$ \\ 
Kepler-351 & $-0.13\pm0.5$ & \tnote{t} & $0.8668\pm0.029$ & $3\pm4.5$ \\ 
Kepler-419 & $0.17\pm0.04$ & \tnote{m} & $0.9932\pm0.013$ & $45.3\pm7.5$ \\ 
Kepler-424 & $0.42\pm0.04$ & \tnote{m} & $1.424\pm0.015$ & $42\pm10.4$ \\ 
Kepler-442 & $-0.37\pm0.1$ & \tnote{u} & $2.727\pm0.017$ & $126.3\pm9.75$ \\ 
Kepler-443 & $0.02\pm0.04$ & \tnote{m} & $1.234\pm0.029$ & $81.4\pm21.6$ \\  
Kepler-452 & $0.21\pm0.09$ & \tnote{v} & $1.805\pm0.01$ & $147.9\pm3.1$ \\ 
Kepler-51 & $0.05\pm0.04$ & \tnote{m} & $1.246\pm0.017$ & $62.4\pm7.65$ \\ 
Kepler-553 & $0.18\pm0.04$ & \tnote{m} & $1.341\pm0.019$ & $105.6\pm15.7$ \\ 
Kepler-56 & $0.4\pm0.04$ & \tnote{m} & $1.076\pm0.012$ & $0\pm0$ \\ 
Kepler-62 & $-0.34\pm0.04$ & \tnote{m} & $3.321\pm0.011$ & $0\pm0$ \\ 
Kepler-712 & $-0.08\pm0.04$ & \tnote{m} & $1.083\pm0.031$ & $136.9\pm13.4$ \\ 
Kepler-87 & $-0.17\pm0.03$ & \tnote{w} & $0.7803\pm0.015$ & $185.4\pm12.5$ \\ 
Kepler-967 & $-0.03\pm0.11$ & \tnote{l} & $2.1\pm0.01$ & $145.8\pm36.5$ \\ 
Kepler-97 & $0.03\pm0.06$ & \tnote{b} & $2.502\pm0.016$ & $2.3\pm2.55$ \\ 
KIC 10525077 & $-0.04\pm0.26$ & \tnote{x} & $0.6661\pm0.022$ & $194\pm65$ \\ 
KIC 3526061 & $0.12\pm0.11$ & \tnote{y} & $2.504\pm0.012$ & $170.4\pm1.7$ \\ 
PH2 & $0.017\pm0.0475$ & \tnote{z} & $2.913\pm0.011$ & $2.5\pm3.4$ \\ 
TOI-1288 & $0.07\pm0.09$ & \tnote{aa} & $8.72\pm0.013$ & $50.1\pm12.1$ \\ 
TOI-1669 & $0.26\pm0.06$ & \tnote{ab} & $8.948\pm0.012$ & $52.9\pm19$ \\ 
TOI-1736 & $0.15\pm0.06$ & \tnote{ac} & $11.34\pm0.015$ & $61.7\pm2.35$ \\ 
TOI-199 & $0.22\pm0.03$ & \tnote{ad} & $9.83\pm0.011$ & $45.2\pm5.35$ \\ 
TOI-2134 & $0.12\pm0.02$ & \tnote{ae} & $44.11\pm0.014$ & $0.1\pm0.15$ \\ 
TOI-2295 & $0.316\pm0.04$ & \tnote{af} & $7.946\pm0.01$ & $9.2\pm5.95$ \\ 
TOI-4600 & $0.16\pm0.08$ & \tnote{ag} & $4.62\pm0.011$ & $23.4\pm8.25$ \\ 
TOI-712 & $-0.166\pm0.061$ & \tnote{d} & $17.04\pm0.0096$ & $26.1\pm33.7$ \\ 
WASP-41 & $0.04\pm0.01$ & \tnote{b} & $6.119\pm0.02$ & $74\pm5.15$ \\
WASP-47 & $0.41\pm0.04$ & \tnote{b} & $3.701\pm0.02$ & $29.1\pm6.5$ 
\\

\bottomrule
\insertTableNotes

\end{longtable*}
\end{ThreePartTable}